\documentclass[fleqn,usenatbib]{mnras}
\usepackage{newtxtext,newtxmath}
\usepackage[T1]{fontenc}
\DeclareRobustCommand{\VAN}[3]{#2}
\let\VANthebibliography\thebibliography
\def\thebibliography{\DeclareRobustCommand{\VAN}[3]{##3}\VANthebibliography}

\usepackage{graphicx}	% Including figure files
\usepackage{amsmath}	% Advanced maths commands
\newcommand\Msun{$M_{\odot}$} % \newcommand\Msun{$\mathrm{M_{\odot}}$}
\newcommand\RJ{$R_{\rm J}$} %\newcommand\RJ{$\mathrm{R_{\rm J}}$}

\title[WD Exoplanets with JWST]{A New Method for Finding Nearby White Dwarf Exoplanets and Detecting Biosignatures}

\author[Limbach et al.]{
Mary Anne Limbach,$^{1}$\thanks{E-mail: maryannelimbach@gmail.com}
Andrew Vanderburg,$^{2}$ 
Kevin B. Stevenson,$^{3}$ 
Simon Blouin,$^{4}$ 
Caroline Morley,$^{5}$
\newauthor
\hspace{1mm}Jacob Lustig-Yaeger,$^{3}$ 
Melinda Soares-Furtado$^{6,7}$ 
and Markus Janson$^{8}$
\\
$^{1}$Department of Physics and Astronomy, Texas A\&M University, 4242 TAMU, College Station, TX 77843-4242 USA\\
$^{2}$Department of Physics and Kavli Institute for Astrophysics and Space Research, Massachusetts Institute of Technology, Cambridge, MA 02139, USA\\
$^{3}$Johns Hopkins APL, 11100 Johns Hopkins Rd, Laurel, MD 20723, USA\\
$^{4}$Department of Physics and Astronomy, University of Victoria, Victoria, BC V8W 2Y2, Canada\\
$^{5}$Department of Astronomy, University of Texas at Austin, Austin, TX, USA\\
$^{6}$Department of Astronomy,  University of Wisconsin-Madison, 475 N.~Charter St., Madison, WI 53703, USA\\
$^{7}$NASA Hubble Science Fellow\\
$^{8}$Department of Astronomy, Stockholm University, Stockholm, Sweden
}

\date{Accepted XXX. Received YYY; in original form ZZZ}

\pubyear{2022}

\begin{document}
\label{firstpage}
\pagerange{\pageref{firstpage}--\pageref{lastpage}}
\maketitle

\begin{abstract}
We demonstrate that the {\it James Webb Space Telescope} ({JWST}) can detect infrared (IR) excess from the blended light spectral energy distribution of spatially unresolved terrestrial exoplanets orbiting nearby white dwarfs.  We find that {JWST} is capable of detecting warm (habitable-zone; $T_{\rm eq}=287$\,K) Earths or super-Earths and hot ($400-1000$\,K) Mercury analogs in the blended light spectrum around the nearest 15 isolated white dwarfs with 10~hrs of integration per target using MIRI's Medium Resolution Spectrograph (MRS). Further, these observations constrain the presence of a CO$_2$-dominated atmosphere on these planets. The technique is nearly insensitive to system inclination, and thus observation of even a small sample of white dwarfs could place strong limits on the occurrence rates of warm terrestrial exoplanets around white dwarfs in the solar neighborhood. We find that JWST can also detect exceptionally cold ($100-150$\,K) Jupiter-sized exoplanets via MIRI broadband imaging at $\lambda = 21\,\mathrm{\mu m}$ for the 34 nearest ($<13$\,pc) solitary white dwarfs with 2\,hrs of integration time per target. Using IR excess to detect thermal variations with orbital phase or spectral absorption features within the atmosphere, both of which are possible with long-baseline MRS observations, would confirm candidates as actual exoplanets. Assuming an Earth-like atmospheric composition, we find that the detection of the biosignature pair O$_3$+CH$_4$ is possible for {\it all} habitable-zone Earths (within 6.5\,pc; six white dwarf systems) or super-Earths (within 10\,pc; 17 systems) orbiting white dwarfs with only $5-36$\,hrs of integration using MIRI's Low Resolution Spectrometer (LRS).
\end{abstract}


%%%%%%%%%%%%%%%%%%%%%%%%%%%%%%%%%%%%%%%%%%%%%%%%%%

%%%%%%%%%%%%%%%%% BODY OF PAPER %%%%%%%%%%%%%%%%%%

\section{Introduction}

\label{sec:intro}
White dwarf (WD) exoplanetary science is a rapidly evolving subdomain of astrophysical research. 
The scientific breakthroughs in this field offer a valuable opportunity to refine our understanding of comparative exoplanetology, as the overwhelming majority (more than 99.9\%) of confirmed exoplanetary hosts will end their lives as WDs. 
Therefore, observations of such late-stage exoplanetary systems provide critical constraints on the evolution of the star-planet systems that comprise the bulk of our exoplanet census data. {Only a handful of white dwarf exoplanets have been discovered to date. Discoveries of major planets around white dwarfs include WD 0806-661 b, a directly imaged $\approx 8$M$_{\rm Jup}$ gas giant planet orbiting a white dwarf at 2500 AU \citep{2011ApJ...730L...9L}, PSR B1620-26 (AB) b, a 2.5M$_{\rm Jup}$ gas giant orbit pulsar-white dwarf binary at 23 AU \citep{1993ApJ...412L..33T,2003Sci...301..193S}, MOA-2010-BLG-477Lb, a 1.4M$_{\rm Jup}$ gas giant orbiting a white dwarf at 2.8 AU detected via microlensing \citep{2021Natur.598..272B}, WD J0914+1914 b, a spectroscopically discovered debris from a possible evaporating ice giant at 0.07 AU \citep{2019Natur.576...61G}, and WD 1856+534 b, a transiting gas giant planet candidate orbiting a white dwarf at 0.02 AU \citep{Vanderburg_2020}. }
We refer the interested reader to \cite{veras2021planetary} for a more thorough review of the diversity of known WD exoplanetary systems and their accompanying observational properties. 

Previous authors have suggested that many WD systems may have relic planetary systems, either beyond 5\,AU where exoplanets are likely to survive the red giant phase or inward of 5\,AU if there is an exoplanetary system re-genesis or migration post red-giant phase \citep{2002ApJ...572..556D,2010ApJ...722..725Z}. 
However, exoplanets in orbit about WD hosts have proven particularly challenging to detect. 
{More specifically, the two leading exoplanet {detection} methods (the transit and radial velocity (RV) techniques)} have been far less successful at WD exoplanet detection compared to the robust detection yields of their main sequence counterparts \citep{2021ARA&A..59..291Z}.
This dearth of WD exoplanet detections is not without reason. 
For example, a WD's small {radius} (0.8-2\% $\mathrm{R_{\odot}}$) and low luminosity ($\approx0.001\,\mathrm{L_{\odot}}$) makes transit monitoring a challenge.
To date, only a single WD with an exoplanet candidate (WD 1856b) has been discovered via the transit detection technique \citep{Vanderburg_2020}. 
There are also difficulties associated with leveraging the RV technique to unveil WD exoplanets.
This includes the near-featureless WD spectrum paired with the WD's low luminosity \citep{Endl2018}.
Thus far, there have not been any RV detections of exoplanets (candidates or otherwise) in orbit about WD hosts. {Hence, while RV and transit have detected $>$98\% of all known exoplanets around main sequence stars, only 20\% of exoplanets around WDs have been detected using these two methods.}

The very characteristics that make RV and transit detections difficult among WD hosts can be leveraged as valuable assets by other exoplanet detection techniques. 
For example, the WD's near-featureless spectrum is a helpful attribute for exoplanet searches via the detection of infrared (IR) excess, which can be indicative of cool companions, including the presence of debris disks \citep{2005ApJ...632L.115K,2005ApJ...632L.119B,2008ApJ...674..431F,2012ApJ...760...26B}, late-type stellar companions, brown dwarfs, or exoplanets. 
The IR excess technique was used to detect the first brown dwarf-WD system {\citep{1987Natur.330..138Z,2022Natur.602..219C}} and, since then, has been leveraged to identify many more such systems \citep[e.g., ][]{2011MNRAS.417.1210G, 2015ApJ...806L...5X, %2015ApJ...815...26F, 2016MNRAS.459.1415B, 
2019MNRAS.489.3990R, 2020MNRAS.498...12H, lai2021infrared}.

Attempts have been made with the Spitzer Space Telescope \citep{2004ApJS..154....1W} to search for exoplanet-induced IR excess in WD systems \citep{2008ApJ...681.1470F, 2010ApJ...708..411K}.
However, Spitzer's observations did not result in any candidates, as a small sample of WDs were surveyed and these measurements were biased towards a narrow companion mass range (i.e., the detection of young, giant planets near the brown dwarf boundary).
To detect smaller and/or colder exoplanets orbiting WDs via IR excess requires improved sensitivity at longer wavelengths. JWST/MIRI is particularly well suited for the detection of cold exoplanets {\citep{2021jwst.prop.2243M}}. 
Recent work has suggested using JWST to characterize gas giant exoplanet atmospheres using the IR excess technique \citep{2020ApJ...898L..35S, 2021ApJ...921L...4L}.
More specifically, JWST/MIRI is sensetive to IR excess at $\lambda = 21\,\mathrm{\mu m}$ --- a wavelength regime where gas giant planets, even those colder than 200~K, can be brighter than the WD host \citep{2001PASP..113.1227I}.  

Another possible detection method is direct imaging. This technique that benefits from the reduced contrast ratio among these low luminosity hosts, which are $10^2-10^5\times$ fainter than main-sequence stars.
\cite{2002MNRAS.331L..41B} noted that this favorable planet-to-star contrast ratio would offer an advantage for the direct imaging of exoplanets in these systems. JWST/MIRI is capable of directly imaging spatially-resolved cold Jupiters in orbit about WD hosts {\citep{2021jwst.prop.1911M}}.   
Such a detection would be of great value, as direct imaging has resulted in the detection of just one widely-separated gas giant planet (WD 0806b) in orbit about a WD host \citep{2011ApJ...730L...9L}, even though several near-IR, high-contrast, direct-imaging surveys were conducted over the past two decades \citep[e.g., ][]{2009MNRAS.396.2074H, 2011ApJ...732L..34T, 2011ApJ...736...89J, 2015A&A...579L...8X, 2021MNRAS.500.3920B, 2021A&A...652A.121P,2022AJ....163...81L}. 

{Other exoplanet detection techniques also benefit from} WD's predictable, temporally-stable, near-featureless spectra, which make deviations from a simple blackbody easily detectable.
This includes orbital-brightness modulations (phase curves) \citep{2013MNRAS.432L..11L,2014ApJ...793L..43L}, photometric signatures of transient events \citep{2003ApJ...584L..91J,2015Natur.526..546V}, and spectral signatures of WD accretion (predominantly in the form of metal absorption lines) from disassociating and/or evaporating companions (e.g., WD J0914b; \citealt{2019Natur.576...61G}).

{In this paper, we explore the detectability of WD exoplanets primarily from IR excess, but also include some analysis of direct imaging and phase curve detectability.
We describe a novel exoplanet atmospheric characterization technique: the presence of exoplanet-induced IR excess within the system's unresolved spectral energy distribution (SED), including atmospheric features imprinted upon the exoplanet's emission such as CO$_2$ absorption at $\lambda = 15\mu$m.
This technique exploits the added SED flux contribution from the mid IR bright exoplanet companion emission.}

Our paper is organized as follows. In Section~\ref{sec:Motivation}, we discuss our motivation for exploring this new method for detecting exoplanets in orbit about WD hosts, in Section~\ref{sec:sample}, we describe our sample of nearby WD systems and models. 
In Section~\ref{gasGiants}, we discuss how IR excess can be used to detect extremely cold (down to 75\,K) gas-giant exoplanets. 
In Section~\ref{sec:detect}, we conduct a detailed analysis demonstrating how IR excess can be used to detect warm terrestrial exoplanets orbiting WDs.
We then describe our new technique for biosignature detection on these worlds. 
In Section~\ref{sec:summary}, we discuss possible JWST observation programs to make use of our proposed detection technique and compare our detectable exoplanet parameter space to Gaia. In Section~\ref{sec:conc}, we close with a summary of our results.

%%%%%%%%%%%%%%%%%%%%%%%%%%%%%%%%%%%%%%%%%%%%%%%%%%%%%%%%%%%%%%
\section{{Methods}}\label{sec:Motivation}
\begin{figure*}
\centering
\includegraphics[width=0.95\textwidth]{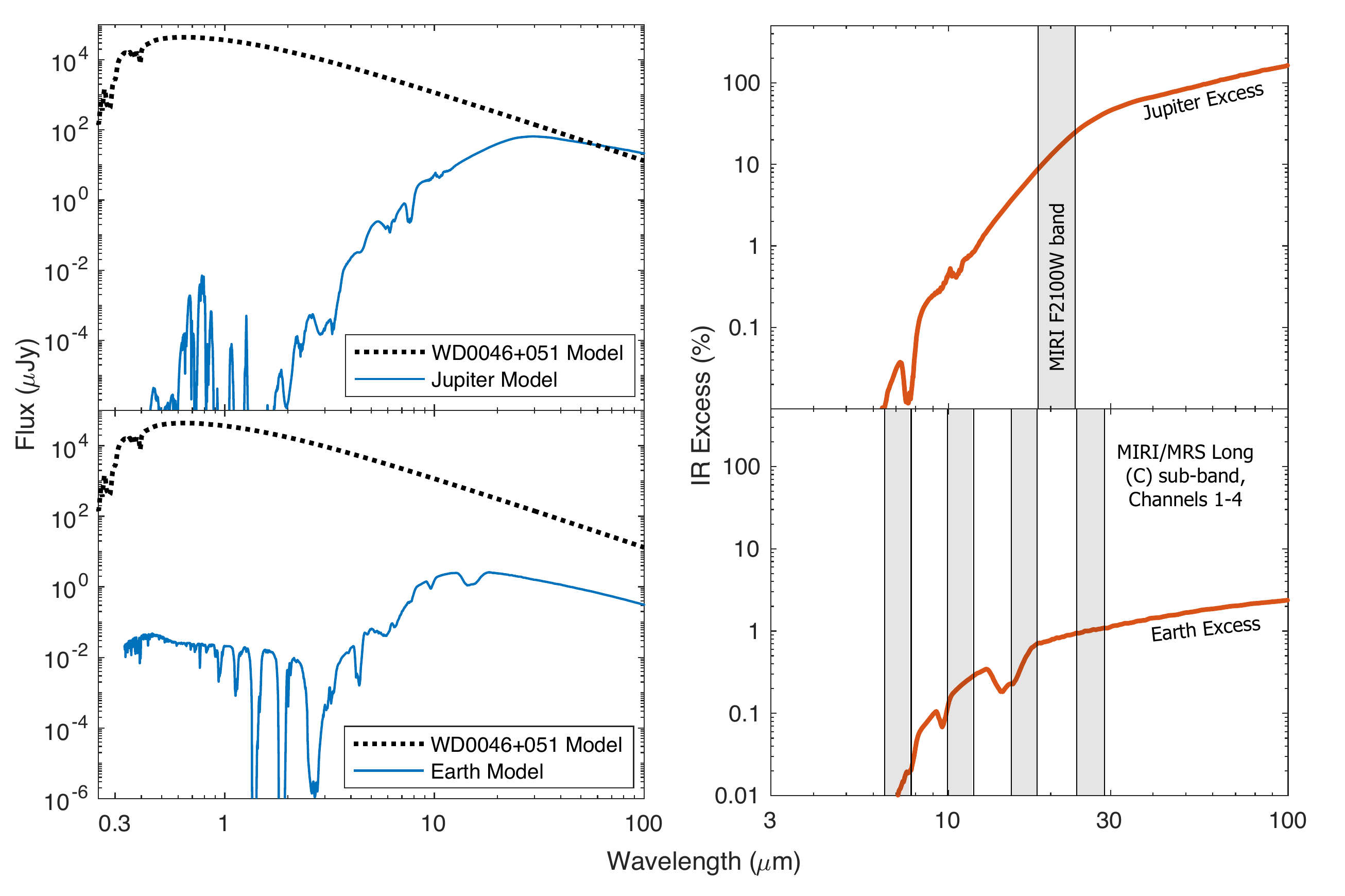}
\caption{{\bf Left panels:}~Flux, in micro-Janskys, from a modeled spectrum of WD~0046+051 ($d$ = 4.32\,pc) compared to a modeled Jupiter-like exoplanet with $T{_{\rm eq}}$ = 150~K, R = $1\,R_{\rm Jup}$, clouds ($f_{\rm sed} = 8$) and no illumination (top) and a modeled Earth-like exoplanet (the model is discussed in Section~\ref{DetectEarths}) with an illumination equivalent to Earth. Despite its cold Jupiter temperature, and because of its comparatively large size, the Jupiter-like exoplanet is brighter than the WD beyond 30\,$\mathrm{\mu m}$. 
{\bf Right panels:}~Percent IR excess from the Jupiter-like exoplanet (top) and Earth-like exoplanet (bottom). At 17~$\mathrm{\mu m}$, the 1~$\mathrm{R_\oplus}$ exoplanet, shown here, is $\sim$0.7\% the flux of the WD. A super-Earth (1.7~$\mathrm{R_\oplus}$) would be $\sim$2\% the brightness of the WD. For comparison, at similar IR wavelengths, the contrast ratio of the Earth-to-Sun is $\sim10^{-6}$. Two example MIRI instrument configurations and bands used in this manuscript are shown by the shaded gray region.} 
\label{IRexcess}
\end{figure*}

In this section, we {describe the methods used to determine the detectability of WD exoplanets with JWST using IR excess and phase curve observations. Our calculations of detectability rely upon (1) models of the white dwarf, (2) models of the exoplanets and (3) JWST/MIRI's sensitivity. The details of our white dwarf models are described in section \ref{sec:sample}, the exoplanet models are described at the beginning of each section in which they are used, and the JWST Exposure Time Calculator (ETC)\footnote{jwst.etc.stsci.edu} and custom code is used to calculate MIRI's sensitivity.

We explore multiple instrument configurations and types of exoplanets throughout the manuscript, but our general process for determining detectability and SNR (signal-to-noise ratio) is as follows. First, we construct modeled SED of the WD of interest (for example, see WD~0046+051, black dotted line, in Figure~\ref{IRexcess}). Then we use existing or create new models of the exoplanet's emission. For example, see the two hypothetical exoplanets analogous to Jupiter (top) and Earth (bottom) in see left panels of Figure~\ref{IRexcess}. Using these models we can then calculate the infrared excess from the exoplanet relative to the WD as a function of wavelength (right panels of Figure~\ref{IRexcess}). We then calculate the SNR in the MIRI configuration of choice (broadband imaging, low resolution spectrograph; LRS, or medium resolution spectrograph; MRS). To do this we compute the combined modeled SED of WD and modeled exoplanet. The combined SED is then uploaded into the JWST ETC where we determined signal-to-noise ratios (SNR). For broadband imaging, this is all that is required to determine SNR. For MRS and LRS, the ETC produces data products including the file ``lineplot\_sn.fits", which gives the SNR of the WD as a function of wavelength. We downloaded this file and developed custom code to bin the SNR of the MRS or LRS measurements to the desired lower resolution spectral bandpasses. Figure~\ref{IRexcess}, right panel, illustrates some of the key MIRI bandpasses that are used in this manuscript. In all configurations, we add a term to account for expected systematic and calibration errors. MIRI's systematic and calibration errors are dependent upon the type of IR excess measurement being made -- and specifically whether or not we are referencing two spectral bands measured simultaneously or sequentially. The source and magnitude of these errors are quantified each time we discuss a new configuration.
 }

From figure~\ref{IRexcess}, we see that at mid-IR wavelengths, WD targets are comparable in brightness to exoplanets. 
This is because WDs have small {radii} (0.8-2\% {$\mathrm{R_{\odot}}$}) and, as such, are faint even when their {effective temperatures} are high (the typical {effective temperature} for a WD {in our solar neighborhood} (within 12 pc) is approximately {$T_{\rm eff}$} = 6000\,K).
Within the blended spectrum of a WD-planet SED, atmospheric absorption lines and blackbody emission from an exoplanet can produce {spectral features that are easily discernible (0.1-10\%) by visual inspection of the theoretical spectrum} (see Figure~\ref{IRexcess}, right panels). 
Figure~\ref{IRexcess} illustrates the SEDs of WD~0046+051 (a solitary WD at $d$ = 4.32\,pc, $R_*$ = 1.2~$\mathrm{R_\oplus}$, {$T_{\rm eff}$} = 6100\,K) compared to a Jupiter-like planet (top panels; $T{_{\rm eq}}$ = 150\,K, R = 1~R$_{\rm Jup}$, no insolation) and an Earth-like planet (bottom panels; $T{_{\rm eq}}$ = 287~K, R = 1~R$_{\oplus}$, Earth-equivalent insolation). {Here, $d$ is the distance to the white dwarf system, $R_*$ is the WD radius, $T_{\rm eff}$ is the effective temperature of the star, $T{_{\rm eq}}$ is the temperature of the exoplanet and R is the radius of the exoplanet. Note that the radius of the white dwarf and terrestrial exoplanets are given in earth-radii ($\mathrm{R_\oplus}$) throughout this manuscript, and the radii of gas giant planets are given in Jupiter radii (R$_{\rm Jup}$).}
In the MIRI 21~$\mathrm{\mu m}$ spectral band, the Jupiter-like planet is 19\% the brightness of the WD. The Earth-like planet contributes a 0.7\% (7000\,ppm) IR excess to the WD's SED on the red end of the MIRI/mid-resolution spectrograph (MRS) $17~\mathrm{\mu m}$ channel. For comparison, in the same spectral band, Earth contributes a 0.0001\% (1\,ppm) IR excess to the Sun. Note that in both cases (for the Jupiter and Earth-like planets) the only features present in the mid-IR spectrum are molecular absorption features in the planets' atmospheres---most notably CO$_2$, but also O$_3$, CH$_4$ and H$_2$O. The only features in the WD spectrum are blueward of 500~nm. The temporal stability and near-featureless spectrum of WDs is key to the successful detection of IR excess and atmospheric features.

%%%%%%%%%%%%%%%%%%%%%%%%%%%%%%%%%%%%%%%%%%%%%%%%%%%%%%%%%%%%%%%%%

\section{The Solar Neighborhood White Dwarf Sample}\label{sec:sample}
For this study, we modeled the SEDs of 34 well-separated or solitary WDs within 13 parsecs of the solar system. 
Table \ref{Table:WDs} gives the sample of WDs used in this study. In addition to several parameters for each WD, the table also includes the minimum radius of an Earth-like planet that can be detected with 10 hours of MIRI observations (column 10), which is discussed in Section~\ref{sec:detect} of this paper, and finally in column 11 we give the equilibrium temperature of a body orbiting at the Roche limit in all 34 systems. 

\begin{table*}
	\centering
	\caption{Solitary (or well separated) WDs within 13 parsecs of the solar system.}
	\label{Table:WDs}
	\begin{tabular}{cccccccccccc}
\hline
\hline
{WD} & {Common} & {G$^b$} &{} & {D} &{$T_{\rm eff}$} & {Mass} & {Radius} & {Age} & {Min. Earth} & {$T_{\rm eq}$~(K) at} & \\
{Number$^a$} & {Name} & {(mag)} & {SpT} & {(pc)} & {(K)} & {(\Msun{})} & {(R$_{\earth}$)$^c$} & {(Gyr)} &  {(R$_{\earth}$)$^d$} & {Roche L.} & {Refs}
\\
\hline
0642-166&Sirius B&8.52&DA&2.67&25967&0.98&0.93&0.228&0.67&1248&3\\
0046+051&Wolf 28&12.30&DZ&4.32&6106&0.70&1.20&3.3&0.83&354 & 1\\
1142-645&GJ 440&11.42&DQ&4.64&7951&0.58&1.37&1.29&0.91&506 & 1\\
1748+708&GJ 1221&13.78&DXP&6.21&5177&0.63&1.29&5.86&1.09&316&5\\
0552-041& HL 4&14.24	&DZ&6.44&4430&0.55&1.39&7.89&1.14&287&2\\
0553+053&V* V1201 Ori&13.97&DAH&7.99&5785&0.72&1.20&4.25&1.41&333&3\\
0752-676&LAWD 26&13.78&DA&8.17&5638&0.55&1.43&4.5&1.46&370&6\\
1334+039&Wolf 489&14.37&DA&8.24&4971&0.54&1.43&5.02&1.46&328&3\\
2359-434&LAWD 96&12.90&DA&8.34&8468&0.83&1.08&1.37&1.48&452&6\\
J2151+5917&Gaia DR2...&14.37&DAH&8.46&5095&0.57&1.39&5&1.49&329&5\\
0839-327&CD-32 5613&11.82&DA&8.52&9203&0.49&1.59&0.55&1.58&652&6\\
2251-070&GJ 1276&15.42&DZ&8.54&4170&0.61&1.31&8.48&1.48&258&2\\
0038-226&GJ 2012&14.30&DQpec&9.10&5210&0.51&1.44&4.44&1.59&348&2\\
0738-172&GJ 283 A&12.97&DZA&9.16&7545&0.57&1.38&1.44&1.57&485&1\\
0435-088&GJ 3306&13.60&DQ&9.41&6395&0.55&1.40&1.85&1.65&415&2\\
0141-675&GJ 3112&GJ 31&DA&9.72&6362&0.58&1.40&1.56&1.70&411&6\\
1917-077&LAWD 74&12.25&DBQA&10.10&10396&0.62&1.32&0.65&1.77&644&3\\
0912+536&EGGR 250&13.78&DCP&10.28&7170&0.74&1.15&2.45&1.77&401&2\\
0310-688&CPD-69 177&11.41&DA&10.40&16444&0.68&1.31&0.16&1.88&998&6\\
1202-232&WD 1202-232&12.74&DAZ&10.43&8667&0.59&1.41&0.83&1.85&558&6\\
0821-669&WD 0821-669&15.08&DA&10.67&4808&0.53&1.45&6.58&1.84&321&6\\
0009+501&GJ 1004&14.24&DAH&10.87&6445&0.75&1.18&3.02&1.87&365&2\\
0245+541&GJ 3182&15.12&DAZ&10.87&4980&0.62&1.32&7.24&1.88&309&2\\
1647+591&V* DN Dra&12.28&DA&10.94&12370&0.79&1.14&0.45&1.92&682&5\\
2140+207&GJ 836.5&13.17&DQ&11.03&7515&0.50&1.49&0.82&1.91&513&1\\
0727+482A&EGGR 52&15.06&DA&11.10&4934&0.51&1.48&4.68&1.91&334&3\\
0727+482B&EGGR 52&15.25&DA&11.10&4926&0.65&1.28&7.13&1.90&298&3\\
0548-001&GJ 1086&14.43&DQ&11.21&6080&0.66&1.24&3.63&1.92&361&2\\
1953-011&GJ 772&13.59&DA&11.56&7752&0.70&1.24&1.63&1.99&455&4\\
1345+238&GJ 1179 B&15.34&DA&11.87&4605&0.45&1.56&4.72&2.05&327&2\\
1917+386&GJ 1234&14.47&DC&11.87&6140&0.65&1.26&3.55&2.04&368&2\\
1055-072&LAWD 34&14.24&DC&12.28&7155&0.77&1.11&2.93&2.09&392&2\\
1900+705&GJ 742&13.25&DAP&12.88&12144&1.01&0.87&0.91&2.19&562&5\\
1620-391&CD-38 10980&11.00&DA&12.91&26209&0.69&1.35&0.02&2.33&1608&6\\
\hline
\end{tabular}
\\{{(a) }{The white dwarf number is given by the method described in \cite{1987ApJS...65..603M}.}}
{(b) }{Gaia magnitudes from EDR3 (\citealt{2021MNRAS.508.3877G}; \url{gea.esac.esa.int/archive/}).}
{(c) }{The radius of the WD is given in earth radii to allow for comparison of the host-size with the exoplanets we are attempting to detect.}
{(d) }{Minimum detectable size of an Earth-like planet at $T_{\rm eq}$ = 287\,K orbiting each WD using JWST/MIRI/MRS/Long Band C, 10 hours of observation time.}
{\it References:} 
[1]	\cite{2019ApJ...885...74C},
[2] \cite{2019ApJ...878...63B},
[3] \cite{2012ApJS..199...29G},
[4] \cite{2020ApJ...898...84K},
[5] \cite{2020MNRAS.499.1890M},
[6] \cite{2019MNRAS.482.4570G}
\end{table*}
%\end{longrotatetable}

The SEDs of the objects listed in Table~\ref{Table:WDs} are modeled using an updated version of the model atmosphere code described in \cite{blouin2018a,blouin2018b} and references therein. Until now, the need for the inclusion of spectral features in the IR had not been felt, and most IR absorption features were omitted from model calculations with the notable exception of collision-induced absorption \citep[]{bergeron2002,kowalski2014,blouin2017}. However, the observations proposed in this work require a high-fidelity modeling of the IR spectra, and the model atmosphere code has now been modified to include all relevant spectral lines and molecular bands up to $30\,\mu$m. More precisely,
\begin{itemize}
    \item For DA WDs, {the atmosphere code} previously included $\textrm{H}\,\scriptstyle\mathrm{I}$ lines from the Lyman, Balmer, Paschen, and Brackett series only. {It} now also considers higher series so that all $\textrm{H}\,\scriptstyle\mathrm{I}$ lines up to $30\,\mu$m are included. Those higher series are included using the Vienna Atomic Line Database (VALD, \citealt{VALD}).
    \item For DZ WDs, {the new models} include the transitions from all elements from C to Cu all the way to 30\,$\mathrm{\mu}$m (up from 1\,$\mathrm{\mu}$m previously).
    \item For DQ WDs, {the new models} now include the opacity of C$_2$ molecular bands in the IR, namely the Ballik--Ramsay and Phillips systems. This is done using the Kurucz linelists\footnote{\url{http://kurucz.harvard.edu/linelists.html}}.
\end{itemize}

\begin{figure}
\centering
\includegraphics[width=0.5\textwidth]{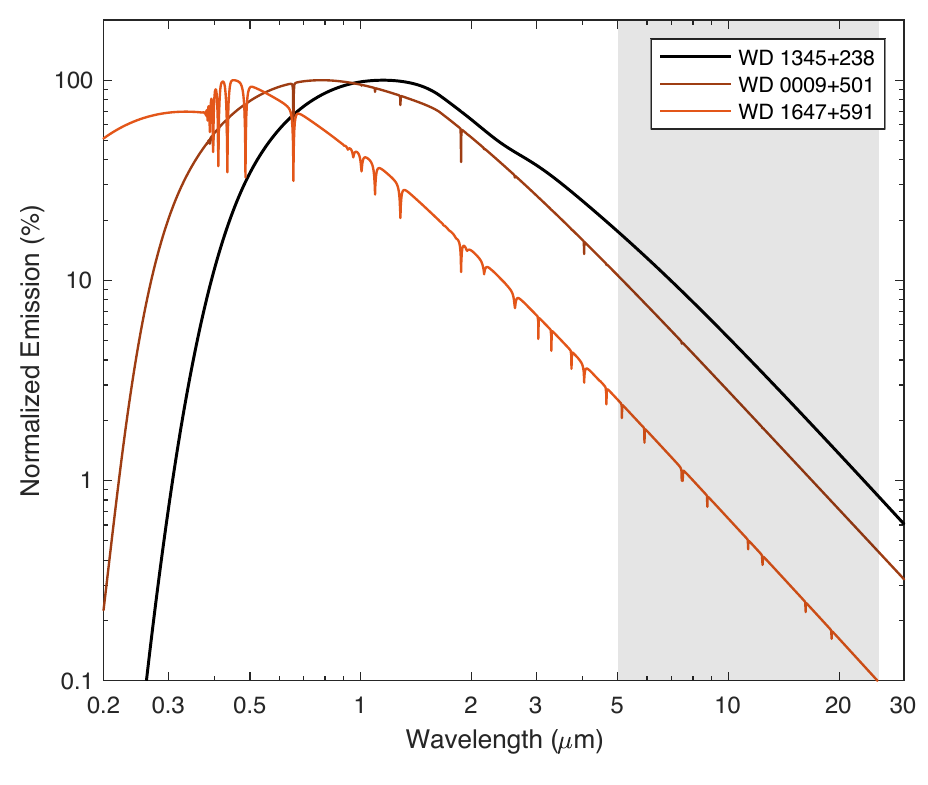}
\caption{Normalized spectral energy distribution (SED) of three WDs in our sample, chosen to illustrate the range of spectral features expected. The three white dwarfs are (from black line to orange line) {$T_{\rm eff}$} = 4600\,K, 6400\,K and 12,400\,K. Although some WDs have a significant number of spectral features in the UV/visible/near-IR, almost all WDs exhibit a featureless Rayleigh-Jeans blackbody tail in the mid- to far-IR, where we propose exoplanet detection with JWST MIRI. The MIRI spectral range is shown in gray. 
}
\label{WDseds}
\end{figure}

Our model SEDs are computed at the nominal parameters given in Table~\ref{Table:WDs}. For metal-polluted atmospheres, metals are included in the model calculations using the abundances given in the references listed in Table~\ref{Table:WDs}. Note that we assumed a pure He atmosphere for the strongly magnetic WD~1748+708, for which the atmospheric composition remains unknown. Figure~\ref{WDseds} shows the normalized SEDs of three WDs in our sample to illustrate the range of diversity as well as their near-featureless spectra in the mid-IR.

%%%%%%%%%%%%%%%%%%%%%%%%%%%%%%%%%%%%%%%%%%%%%%%%%%%%%%%%%%%%%%%%%

%%%%%%%%%%%%%%%%%%%%%%%%%%%%%%%%%%%%%%%%%%%%%%%%%%%%%%%%%%%%%%%%%%%%%%
\section{Finding Frigid Gas Giants}\label{gasGiants}
In this section, we demonstrate that JWST/MIRI broadband imaging is optimal for detection of nearby, cold Jupiters (see Figure~\ref{BroadbandJups}), including both those that are resolved (detection by direct imaging) and unresolved (detection from IR excess).

\subsection{Cold Jupiter Models}\label{JupModels}

We generated custom simulated spectra of cold Jupiter analogs (those $<200$\,K) using a 1D atmosphere model, described in detail in \citet{Marley1999, Saumon2008, Morley2012, Morley2014}, with updates to the molecular and atomic opacities as described in \citet{Marley2021}. The models assume that the atmosphere is in radiative--convective and `rainout' chemical equilibrium, in which species condense into clouds as the atmosphere cools. All models have Jupiter-like metallicity, [M/H]=0.5, increasing the abundance of all elements heavier than helium by a factor of 3.16. Surface gravities of all models are 25 m/s$^2$ (same as Jupiter). Clouds made of water ice are included using the method presented in \citet{AM01}, where varying the sedimentation efficiency $f_{\rm sed}$ changes the vertical extent and mode particle sizes of clouds. Ammonia clouds are not included in the calculation. Model atmospheres are calculated for $T{_{\rm eq}}$=75, 100, 125, 150, and 175\,K. Spectra are calculated at moderate resolution using the thermal emission code presented in the appendix of \citet{Morley2015}. 

For simulated spectra from $T{_{\rm eq}}=200-300$\,K, we use the publicly available Sonora-Bobcat model spectra. These spectra are described in \citet{Marley2021}; in short, they are calculated using the same 1D atmosphere code as above \citep{Marley1999, Saumon2008, Morley2012, Morley2014}, also assuming radiative-convective and `rainout' chemical equilibrium. We use models with Jupiter-like metallicity ([M/H]=0.5) and gravity (log g=3.5). 

\subsection{IR Excess}
\begin{figure*}
\centering
\includegraphics[width=0.95\textwidth]{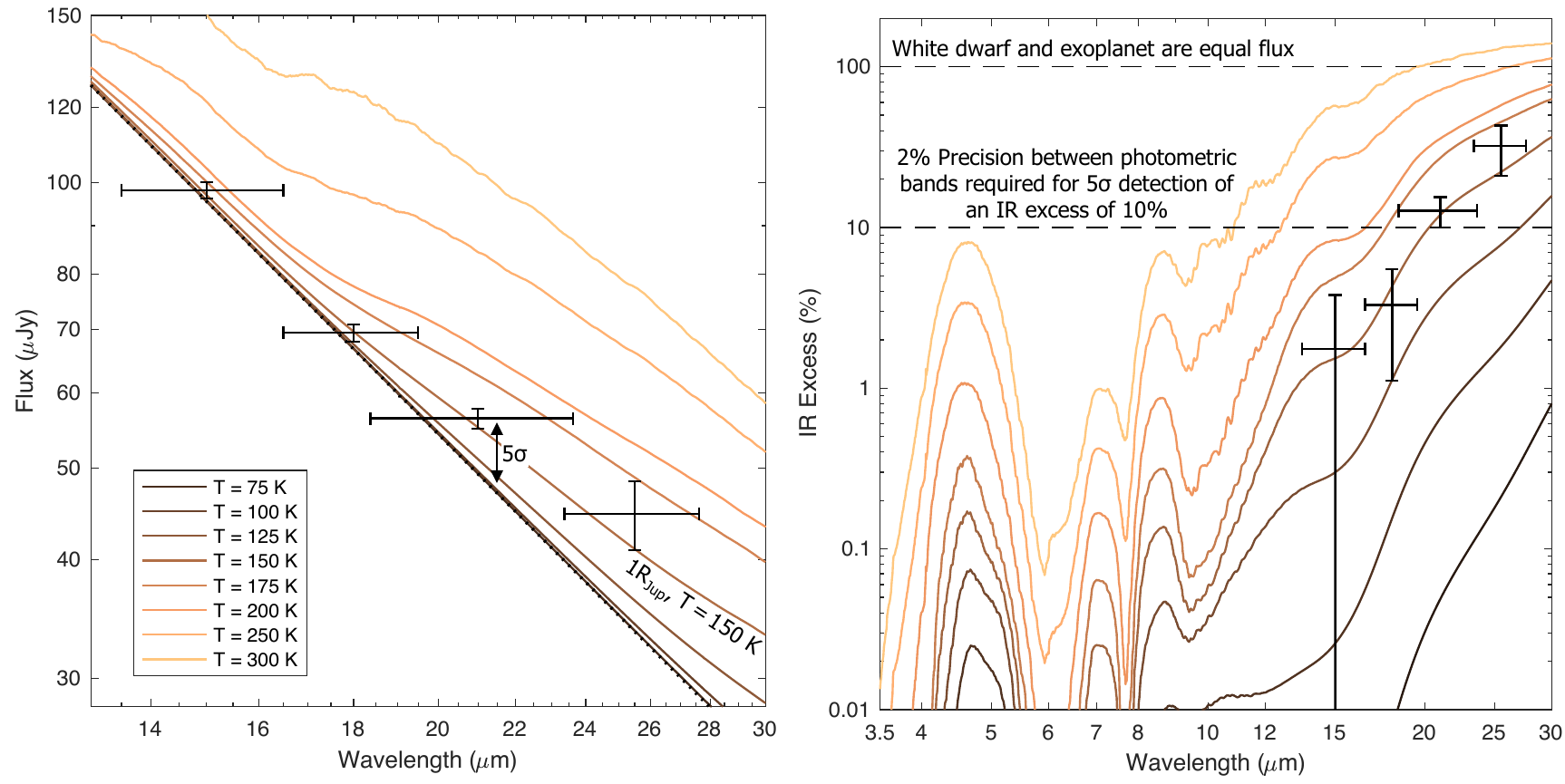}
\caption{Combined flux (left) or IR excess (right) from an unresolved WD and Jupiter-sized exoplanet assuming various planet temperatures ($T{_{\rm eq}}$~$=75-300$\,K). The black crosses are simulated MIRI/broadband imaging data of an unresolved 150~K Jupiter-analog orbiting WD 0009+501 ($d$ = 10.9\,pc) assuming 2 hrs of observation per band. The 1$\sigma$ detection limits are given by the root sum square of the (1)  photometric error based on the JWST/ETC calculations and (2) assuming a 2\% error \citep{2022arXiv220406500G} referencing between spectral bands (to achieve a 5$\sigma$ detection of the IR excess from the planet requires a minimum excess of 10\%). In this example, the exoplanet is detected in the 21$\mathrm{\mu}$m band at 5$\sigma$. For $T{_{\rm eq}}$$>$150\,K exoplanets at distances $\lesssim12$\,pc, detection (5$\sigma$, 2hrs) is possible at 21\,$\mathrm{\mu}$m.  Detection at wavelengths shorter than 20\,$\mathrm{\mu}$m is precluded by insufficient excess ($<10\%$). Beyond 25\,$\mathrm{\mu}$m detection is inhibited by thermal background and photon noise, and for distances $\gtrsim12$\,pc, this noise source precludes detection at 21\,$\mathrm{\mu}$m as well.} 
\label{BroadbandJups}
\end{figure*}

{In this section, we will determine the amount of IR excess that is detectable via MIRI broadband imaging.} The IR excess from cold Jupiters is often a large fraction ($\gtrsim$10\%) of the combined WD + exoplanet system flux at 21\,$\mathrm{\mu}$m. {To constitute a large fraction of the system flux requires that the exoplanet is sufficiently massive and/or young. Specifically, to reach 10\% excess at 21\,$\mathrm{\mu}$m, requires that the exoplanet is $T{_{\rm eq}}$~$>$\,150\,K, which corresponds to a planet mass $>$1M${\rm _{Jup}}$ for systems with an age $<$5\,Gyr or M${\rm _{planet}}>$1M{$\rm _{Saturn}$} for systems that are $<$1.5Gyr (see Figure \ref{BroadbandJups}; \citealt{2019A&A...623A..85L}).}

Figure~\ref{BroadbandJups} shows the detectability of cold Jupiter-sized planets assuming two hours of observation on the WD 0009+501 ($d$ = 10.9\,pc) system. To calculate the system flux and detectability, we computed the combined modeled SED of WD 0009+501 with modeled exoplanets of various temperatures (75-300\,K) and a common radius of 1\,\RJ{}. The flux of the combined components is shown in the left panel of Figure~\ref{BroadbandJups}, while the IR excess from the exoplanet (relative to the WD) is shown in the right panel. 
The combined SED was then uploaded into the JWST ETC where we determined SNR in all MIRI imaging spectral bands as well as the NIRCam $4.5\,\mathrm{\mu}$m band using 2\,hrs of total integration time.
The resulting simulated JWST/MIRI/broadband data is illustrated for a single 150\,K exoplanet (black points with error bars). The photometric precision is shown by the vertical error bars in the figure, while the horizontal bars illustrate the width of each spectral band.

JWST 21\,$\mathrm{\mu}$m broadband imaging would result in a 5$\sigma$ detection of a 150~K Jupiter in only 2 hours. While we examined exoplanet detectability in other bands, we found 21\,$\mathrm{\mu}$m to be optimal. At wavelengths shorter than 21\,$\mathrm{\mu}$m, there is insufficient IR excess in the unresolved planet-WD SED, which would make referencing between spectral bands difficult. Specifically, for our calculations, we assume we can achieve a 2\% precision {(justified below)} when referencing between two bands that are not imaged simultaneously. This corresponds to a minimum detectable IR excess of 10\% assuming a 5$\sigma$ detection is desired. At longer wavelengths, there is a larger percentage of IR excess from the exoplanet, but JWST does not possess sufficient photometric precision (limited by photon noise/thermal background) to detect the planet with high confidence. Therefore, the detection of gas-giant exoplanets around the {\it nearest} WDs is most efficient at 21\,$\mathrm{\mu}$m. 
If WD 1856b (a cool gas giant exoplanet candidate orbiting a WD at 25\,pc) was within 15\,pc, it would be detectable in $<3$ hours of JWST 21\,$\mathrm{\mu m}$ even if its temperature is only $T{_{\rm eq}}=150$\,K.

For systems within 25\,pc, JWST/MIRI 21\,$\mathrm{\mu m}$ broadband direct imaging is able to detect even colder gas giants if they are spatially resolved from the host WD. In such cases, the $>$10\% IR excess for detection is unnecessary and only the exoplanet needs to be detectable.
With this approach, with three hours of observation time, MIRI is capable of detecting (5$\sigma$) Jupiter-sized planets as cold as 90\,K at $d<4$\,pc and 125K at $d<8$\,pc.
Spatially resolved or unresolved, Jupiters that are 150\,K are detectable out to about 15\,pc, beyond which the thermal background prohibits detection for observation times shorter than a few hours.  

MIRI has a nominal requirement of 2\% absolute flux prediction accuracy of standard stars\footnote{\url{jwst-docs.stsci.edu/jwst-data-calibration-considerations/jwst-data-absolute-flux-calibration}}\citep{2022arXiv220406500G}. {We note that the flux calibration of HST instruments was based on three DA white dwarf stars. The calibrations of the HST and IRAC were also studied using A-dwarfs, solar analogs, and white dwarfs and found to be consistent within 2\% \citep{2011AJ....141..173B}. Because of their stability, white dwarfs are often chosen as standard stars, and 2\% precision should be achievable for all the white dwarfs in our sample.}
Therefore, we assume it will be possible to achieve measurements between spectral bands of our WD targets at the 2\% photometric precision level. 
IR excess can be detected by measuring the flux at two different wavelengths---a reference band (for example, 5\,$\mathrm{\mu}$m) and the 21\,$\mathrm{\mu}$m band (i.e., the MIRI broadband F2100W imaging filter). 
A WD SED model can be fit to the shorter wavelength measurement; an excess above the model fit at 21\,$\mathrm{\mu}$m may be indicative of an exoplanet.
Such an exoplanet candidate can be flagged for follow-up. 
%Among all nearby WD systems, Jupiters as cold as 150\,K have sufficient IR excess to permit detection within $\leq 2$\,hrs. 

\begin{figure}
\centering
\includegraphics[width=0.5\textwidth]{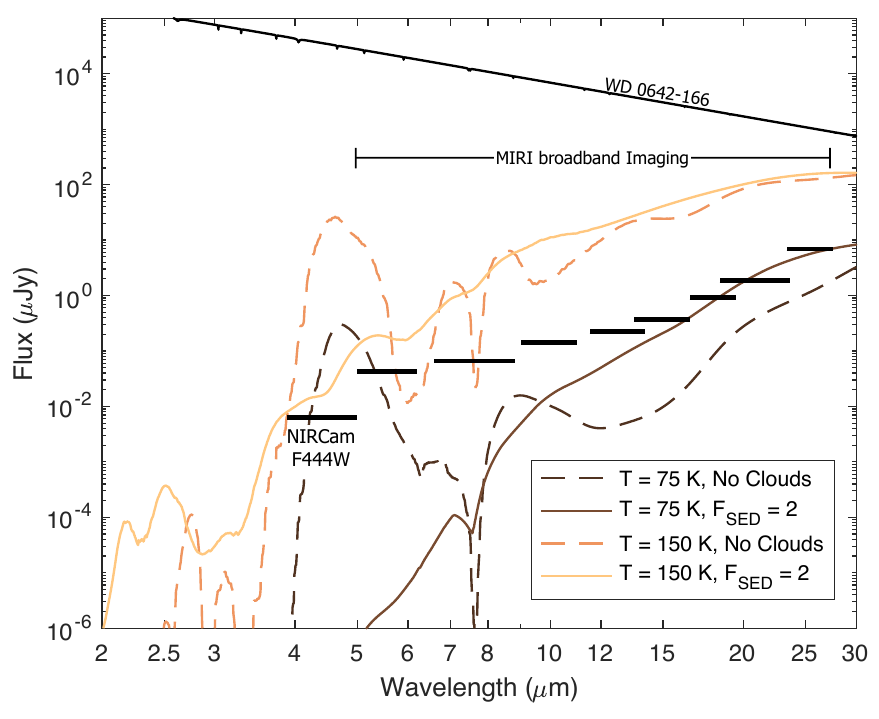}
\caption{Flux from four Jupiter-sized exoplanets at two temperatures and with varying amounts of cloudiness (orange and brown lines) at a distance of 2.67\,pc. Black bars indicate the detection limits (10\,hrs, 5$\sigma$) of NIRCam/F444W (lower left bar) and MIRI/broadband imaging (all other bars). A 150\,K exoplanet is detectable in all bands regardless of cloudiness. 
Conversely, the cloudy, 75\,K planet is only detectable with MIRI 21\,$\mathrm{\mu}$m band imaging, whereas the clear 75\,K planet is only detectable with NIRCam/F444W. The MIRI 21\,$\mathrm{\mu}$m band is optimal for detecting IR excess or direct imaging of cold, cloudy planets.} 
\label{Clouds}
\end{figure}

Figure~\ref{Clouds} illustrates the detectability of 75\,K and 150\,K gas giant exoplanets orbiting WD 0642-166 (Sirius B). 
We depict exoplanets with both cloudy (f$_{\rm sed} = 2$) and clear atmospheres (see description in Section~\ref{JupModels}). 
Many exoplanets models use clear atmospheres, and the detectability of gas giants is usually determined based on these clear models.
It is interesting to note that clear atmosphere models suggest cold gas giants are detectable at $4-5$\,$\mathrm{\mu}$m.
Yet, if these planets are partly cloudy, detection is much easier at longer wavelengths and most efficient with JWST at 21\,$\mathrm{\mu m}$. Indeed, one might expect that all gas-giants possess some clouds, which is perhaps why previous direct imaging surveys \citep[e.g., ][]{2009MNRAS.396.2074H, 2011ApJ...732L..34T, 2011ApJ...736...89J, 2015A&A...579L...8X, 2021MNRAS.500.3920B, 2021A&A...652A.121P,2022AJ....163...81L} for gas-giants orbiting WDs at $4-5$\,$\mathrm{\mu}$m did not detect any companions.

\subsection{Phase Curves}
Using two-photometric-band IR excess, it is possible to detect WD exoplanet false positives. { Fortunately, there are a couple techniques for confirming exoplanets identified via IR excess which should make it possible to eliminate false positives. First, we note that nearby WDs have very high proper motion, and background objects will be stationary, so if the system is observed a few months later, there will no longer be a chance alignment. With large proper motions, past WD observation from HST or the ground should confirm the lack of a background star contaminating the signal. Further, MIRI's high spatial resolution (0.11 arcsec/pixel) gives a low probability of contamination.}
Second, while false positives from both debris disk and unresolved background stars are capable of producing IR excess, neither would produce a phase curve. 
This makes phase curves a good method to rule out non-planetary false positives.

\begin{figure}
\centering
\includegraphics[width=0.5\textwidth]{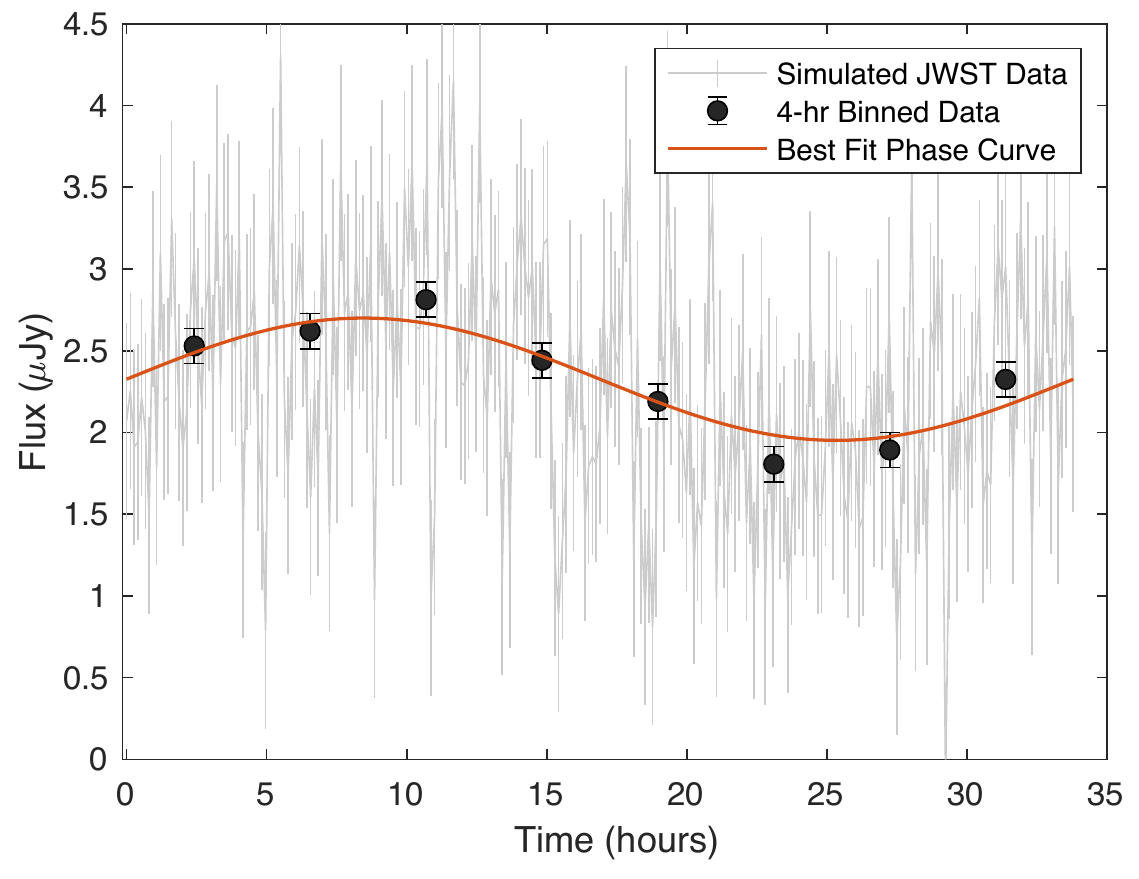}
\caption{Simulated JWST/MIRI/15\,$\mathrm{\mu}$m lightcurve of WD 1856b (assuming the object has a day-side temperature of 225\,K, night-side 150\,K). Light gray data points correspond to each individual 8-min integration, and black data points are the 4-hr binned median of the data. The measured flux is the combined $15\,\mathrm{\mu}$m flux of the WD + exoplanet. If the exoplanet has a day-side temperature of 225\,K, it constitutes a large fraction, {up to 46\% of the total system flux at conjunction.}}
\label{PhaseCurve}
\end{figure}

Figure~\ref{PhaseCurve} shows a simulated phase curve for WD 1856b {(0.02 AU planet-WD orbital semimajor axis, 1.4 day orbit; \citealt{Vanderburg_2020})} to illustrate its detectability, assuming {the exoplanet is tidally locked with} a day-night temperature difference of 75\,K and a day-side temperature of 225\,K. 
Note that the flux from the entire system is plotted here, and the exoplanet contributes significantly to the total system flux at 15\,$\mathrm{\mu}$m. We note that if the upcoming JWST Cycle 1 transmission spectrum observations of WD 1856b indicate that the planet is warm and cloudy, this would suggest that detecting emission from the planet via phase curve observations at 15\,$\mathrm{\mu}$m could be more successful than eclipse emission detection at $4-5$\,$\mathrm{\mu}$m, as cloudy planets emit farther in the IR (see Figure~\ref{Clouds}).

While the presence of IR excess at 21\,$\mathrm{\mu m}$ can identify exoplanet candidates, confirmation is still required.  Phase curve observations can be used to provide such confirmation. 
JWST phase curve observations would be possible for nearby ($d<15$\,pc) cold gas giants at 21\,$\mathrm{\mu m}$ or at further distances at 15\,$\mathrm{\mu m}$ if the planet is sufficiently bright/hot. 
More specifically, although WD 1856b is too far away for phase curve observations at 21\,$\mathrm{\mu m}$, if the planet is sufficiently warm and has a notable day-night temperature difference, it would be possible to measure the exoplanet's phase curve modulations at 15\,$\mathrm{\mu m}$.

\section{Detecting Terrestrial Exoplanets \& Biosignatures}\label{sec:detect}

The emission from gas-giant exoplanets can be comparable in flux to the WD host {in the mid IR}, but terrestrial worlds are much fainter, and therefore detection of these worlds requires observations with JWST/MRS to enable careful reference between multiple spectral bands that are measured simultaneously. In this section, we establish an observational procedure with JWST to search for terrestrial exoplanets and detect biosignatures in nearby WD systems. 

\subsection{Step 1: Detecting Exoplanets via IR Excess}
\subsubsection{The JWST/MIRI/MRS Detection Approach}
{In this section we will demonstrate} that JWST is sufficiently sensitive to detect IR excess from an Earth-analog, and determine the most efficient method for detection.
We consider the detectability of two types of terrestrial planets: (1) those with an Earth-like atmosphere and (2) rocky planets with no atmosphere. To this end, we use the following process to calculate the detectability of exoplanets from IR excess in planet-hosting WD systems.

We uploaded all the modeled WD spectra from our sample (described in Section~\ref{sec:sample}) into the JWST/ETC to perform calculations. Note that the planet contributes negligibly to the total flux of the combined system and was therefore not included in the SED (i.e. although detectable, including the flux from terrestrial exoplanets does not substantially increase the total SNR of the system---this was not the case for gas giant exoplanets).

For several of our targets, we calculated the SNR using several MIRI configurations, including the low resolution spectrograph (LRS) and the MRS A, B and C sub-bands. We found MIRI/MRS sub-band C consistently provided the best SNR on terrestrial exoplanets, and therefore we choose to use this mode for terrestrial exoplanet IR excess detection. The MRS sub-band C spectral bandpasses are illustrated in Figure~\ref{IRexcess}.

Using the JWST ETC\footnote{\url{jwst.etc.stsci.edu}} we calculated the SNR using MRS sub-band C with a total observation time of 10 hours per system while maintaining a $<50\%$ detector saturation at all wavelengths per integration. In all cases, the full subarray was read out in slow readout mode. We required integration times of $\leq$10~min such that the temporal resolution was sufficient for phase curve construction (see Section~\ref{sec:step2}). We computed SNRs using both the JWST ETC MIRI/MRS Imaging and MIRI/MRS time-series modes. The time-series mode gave an SNR $\approx$25\% higher than the MIRI/MRS Imaging mode (for the same total exposure time). However, the JWST ETC overestimates SNR in the time-series mode, so we adopted the conservative SNRs that were based on the MIRI/MRS imaging mode ETC calculations for this manuscript\footnote{\url{jwst-docs.stsci.edu/jwst-exposure-time-calculator-overview/jwst-etc-calculations-page-overview/jwst-etc-time-series-modes}}.
  
The ETC produces data products including the file ``lineplot\_sn.fits", which gives the SNR of the WD as a function of wavelength. We downloaded this file and developed custom code to bin the SNR of the MRS measurements to determine the total SNR from all the photons collected in each of the 4 channels in the MRS C sub-band. These four channels are measured simultaneously with the MRS instrument and correspond to the following wavelengths ranges: 6.4–7.5~$\mathrm{\mu m}$, 10.0–11.8~$\mathrm{\mu m}$, 15.4–18.1~$\mathrm{\mu m}$ and 23.9–28.3~$\mathrm{\mu m}$. We used spectrally-averaged SNRs to calculate the photometric precisions in each channel.
 
We then calculated the amount of IR excess (in parts per thousand) from the terrestrial planet of interest. Finally, the IR excess divided by the photometric precision results in the SNR of our detection of the exoplanet in each MRS channel.

\subsubsection{{WD Systems with Stable Habitable Zones}}\label{RRlimits}

{The habitable zone (HZ) of most WDs in the solar neighborhood lies close to the Roche limit, the minimum stable orbital distance of a secondary within which the tidal forces from the star disintegrate the planet. We are interested in determining the detectability of HZ exoplanets, however, before we do so it is necessary to constrain which WD systems are capable of hosting HZ planets without tidal disruption.} To do this, we must ensure that the HZ lies outside the Roche limit. The Roche limit, $a_{\rm min}$, is given by 
\begin{equation}
    a_{\rm min} = 2.44R_{\rm p}\left(\frac{M_*}{M_{\rm p}}\right)^{1/3}
\end{equation}
where $R_{\rm p}$ is the radius of the exoplanet, $M_*$ and $M_{\rm p}$ is the mass of the star and exoplanet, respectively. We can then compute the equilibrium temperature of the exoplanet at the Roche limit, $T_{\rm eq}$, which is given by 
\begin{equation}\label{Eq:Teq}
    T_{\rm eq} = T{_{\rm eff}}\left(1-\alpha\right)^{1/4}\sqrt{\frac{R_*}{2a_{\rm min}}}
\end{equation}
where $a$ is the distance between the WD and planet (assuming e = 0), $\alpha$ is the albedo of the planet, $T{_{\rm eff}}$ is the temperature of the WD, and we assume uniform heat redistribution. The equilibrium temperature of an Earth-density exoplanet orbiting at the Roche limit is given in Table \ref{Table:WDs} for all nearby WD systems. Here we see equilibrium temperatures that range between $T_{\rm eq} = 258-1608$~K for an exoplanet at the Roche limit. Based on the habitable zone boundaries defined in \cite{2013ApJ...765..131K}, the inner habitable zone lies inside the Roche limit for many cooler WDs, but for all systems in our sample, there is a continuous HZ that stable for $>3$\,Gyr \citep{2011ApJ...731L..31A}. We refer the interested reader to \cite{2011ApJ...731L..31A,2012ApJ...757L..15F,2013AsBio..13..279B,2013MNRAS.432L..11L,2018ApJ...862...69K,2020ApJ...894L...6K} for a more thorough discussion of WD HZs and the habitability of white dwarf exoplanets. We also note that most WD HZ exoplanets are likely to experience strong tidal forces and be tidally locked to their host \citep{2017CeMDA.129..509B}.

\subsubsection{Detecting Earth-Analogs}\label{DetectEarths}

To determine the IR excess from the terrestrial planet, we must assume a SED for the object of interest. For Earth-analogs, we use a model from the Virtual Planetary Lab (VPL\footnote{\url{depts.washington.edu/naivpl/content/vpl-spectral-explorer}}) model deck. 
Specifically, we used a photochemically self-consistent model of the modern Earth orbiting a Sun-like G2V star from \citet{2003AsBio...3..689S} using present atmospheric levels of $O_2$. The top-of-atmosphere spectrum was produced using the Spectral Mapping Atmospheric Radiative Transfer model \citep[SMART;][]{1996JGR...101.4595M}. SMART is a plane-parallel, multi-stream, multi-scattering radiative transfer model that uses the discrete ordinate algorithm, DISORT \citep{1988ApOpt..27.2502S}, and is applicable for scattering, absorbing, and emitting atmospheres. The Earth-like spectrum was calculated assuming no clouds and an ocean covered surface albedo/emissivity.  

\begin{figure}
\centering
\includegraphics[width=0.5\textwidth]{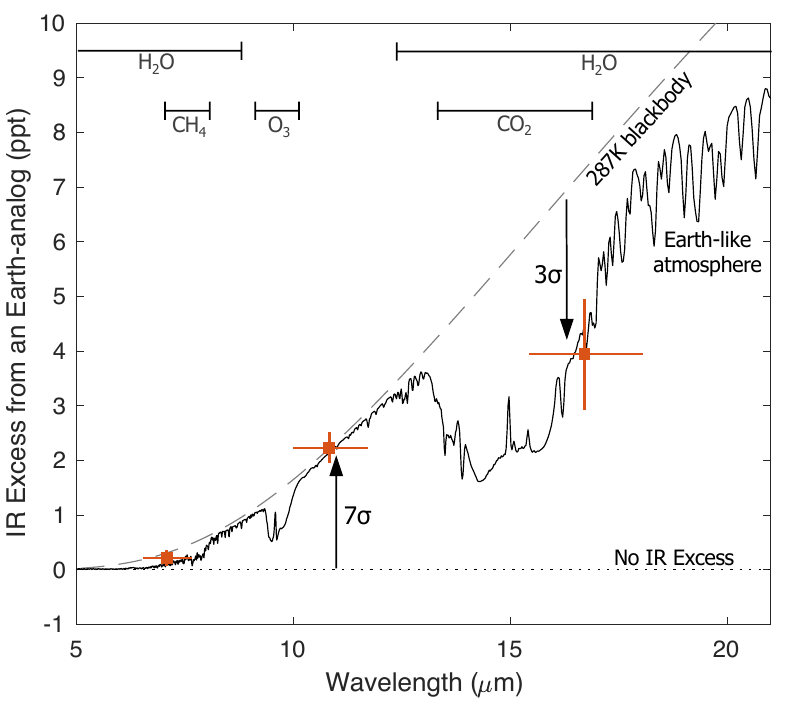}
\caption{IR excess (in ppt) from a 1R$_\oplus$ Earth-analog orbiting WD 0046+051. Simulated JWST data (red crosses) based on 10~hrs of MIRI/MRS sub-band C observations. Data points are binned measurements in each MRS channel. The IR excess from an Earth-analog orbiting WD 0046+051 is detectable (7$\sigma$) in channel 2 ($11\mathrm{\mu m}$). The divergence from a blackbody spectrum (indicative of the presence of an atmosphere and CO$_2$ absorption) is detectable at the 3$\sigma$ level in channel 3. At $17\mathrm{\mu m}$ (channel 3) a habitable-zone Earth is {0.4\%} the flux of its host WD. Note that the horizontal error bars are not statistical uncertainties, but show the width of the spectral band in each channel.}
\label{MRSlimits}
\end{figure}

For habitable-zone (HZ) Earth analogs with similar temperatures and atmospheric compositions, we find that IR excess from the planet is most easily detected (highest SNR) in channel 2 (10.0–11.8~$\mathrm{\mu m}$). {In the next section, we explore follow up observations (with MIRI/LRS) that can characterize the atmosphere of these terrestrial worlds, but we note that even these relatively short MIRI/MRI observations designed to detect terrestrial worlds are capable of detecting the presence of an atmosphere ($CO_2$ and $H_2O$) in channel 3 (15.4–18.1~$\mathrm{\mu m}$) for super-Earths (R $ \approx 1.5 $R$_{\earth}$). }
Figure~\ref{MRSlimits} shows the detectability of a HZ Earth-like exoplanet orbiting WD 0046+051 using the MRS sub-band C observations. We spectrally bin to one point per channel. In Figure~\ref{MRSlimits}, the 1.0~$\mathrm{R_\oplus}$ analog is detectable with a confidence of $7\sigma$ given 10 hours of observations in the MIRI MRS C sub-band, channel 2. Channel 3 also shows a substantial amount of IR excess, adding to the significance of planet detection. If we require instead a $5\sigma$ detection in channel 2, Earth-analogs as small as 0.83~$\mathrm{R_\oplus}$ are detectable in this system (WD 0046+051) with 10\,hrs of observation. Table \ref{Table:WDs} gives the radius of the smallest detectable ($5\sigma$) Earth-analog in all the nearby WD systems using this MIRI/MRS detection approach. The IR excess in Figure~\ref{MRSlimits} from the Earth-analog in channel 1 (6.4-7.5$\mathrm{\mu m}$) is not detectable. However, if this planet were instead a $T{_{\rm eq}}$ = 287~K blackbody (with no atmosphere), there would be a marginally detectable level of IR excess in channel 1. The deficit is due to the presence of methane and water absorption in the exoplanet's atmosphere (the absorption is detectable at the $2.5\sigma$ level). Channel 3 also shows significant deviation from a blackbody spectrum due to CO$_2$ absorption, which is detectable at 3$\sigma$ significance

Due to thermal background, channel 4 (23.9–28.3~$\mathrm{\mu m}$) is very noisy -- except for the nearest few systems, the WD is not detectable, let alone the planet. {While in some cases the planet and WD are detectable at $26\mu m$ for broadband imaging (see Figure~\ref{BroadbandJups}), the MRS slit is large and typical SNRs in channel 4 range between SNR $= 0.1 - 3.0$ due to thermal background. If we were to include the datapoint for channel 4 in Figure 6, the vertical error bars would extend beyond the current range of the y-axis.} Therefore, we leave this channel off our plots (e.g. Figure~\ref{MRSlimits}). We merely note that, to our knowledge, there are no detections of WDs this far in the IR and so if observations with this method were carried out on the nearest systems using JWST, this would provide the first constraints on WD SEDs beyond 24\,$\mathrm{\mu}$m.

\subsubsection{Detecting Rocky Planets with Tenuous Atmospheres}\label{Mercury}

Using equation \ref{Eq:Teq} and $\alpha =  0.14$ gives $T_{\rm eq} = 430$~K for Mercury. From Table \ref{Table:WDs}, 13 of the 34 WDs are capable of hosting exoplanets with an equilibrium temperature equal to or greater than Mercury, and almost all (32/34) are capable of hosting exoplanets with more insolation than Earth. Therefore, it is interesting to explore the detectability of rocky planets with tenuous atmospheres under the assumption that many hot, small exoplanets may not retain any atmosphere due to EUV radiation from the WD \citep{2019ApJ...887L...4S}. For close-in exoplanets with no atmosphere (Mercury analogs) we model the exoplanet with a simple blackbody SED. We then determine which MIRI sub-band C channel is optimal for detection. We find that for cold exoplanets ($\lessapprox 275$~K), detection is optimal in channel 3 (15.4–18.1~$\mathrm{\mu m}$), for warm ($\approx275-650$~K) exoplanets detection is optimal in channel 2 (10.0–11.8~$\mathrm{\mu m}$) and the hottest ($>600$~K) exoplanets are more easily detected in channel 1 (6.4–7.5~$\mathrm{\mu m}$). Shorter wavelength spectral bands are optimum for detection of hotter exoplanets since hot exoplanets emit more flux at shorter wavelengths and JWST MIRI is capable of higher-precision photometry at shorter wavelengths bands.

\begin{figure*}
\centering
\includegraphics[width=1\textwidth]{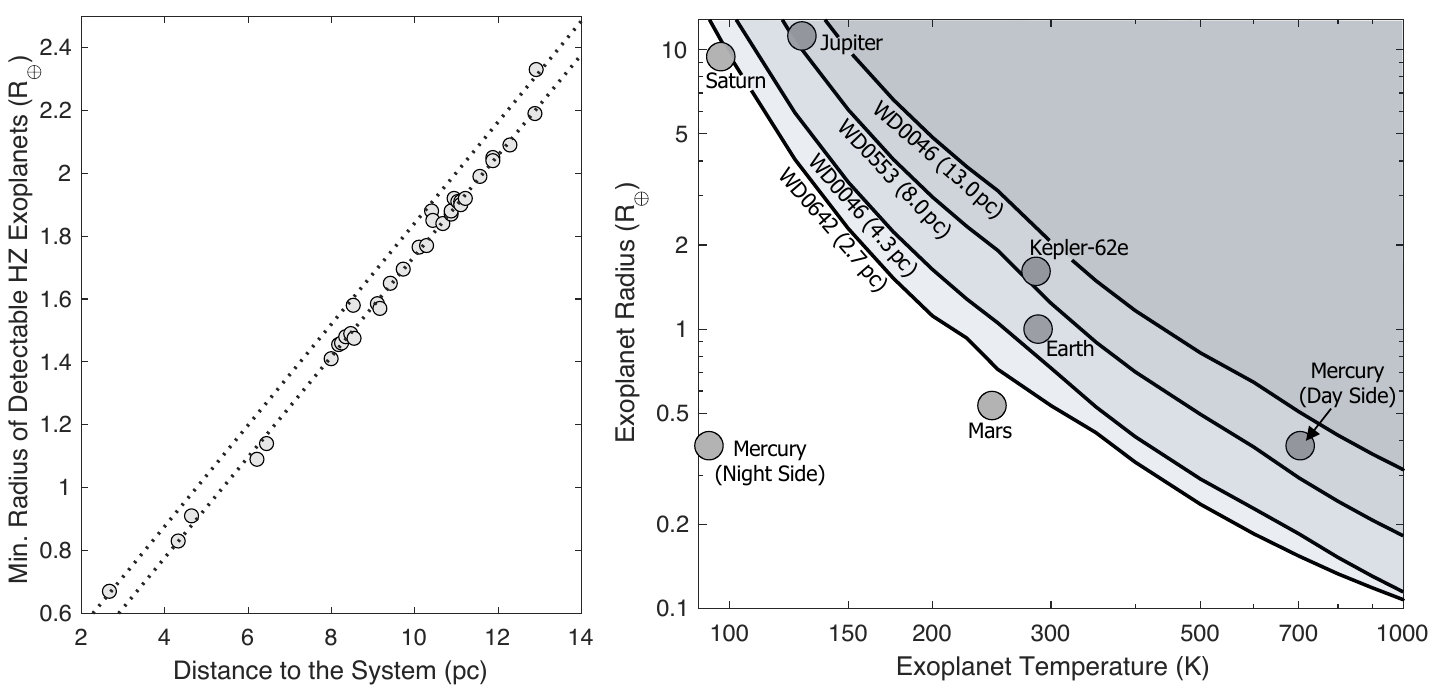}
\caption{{\bf Left:}~Minimum detectable radius of a habitable-zone exoplanet ($T_{\rm eq} = 287$\,K) for each WD in our sample. This figure demonstrates that the size of HZ planet that can be detected depends primarily on the distance to the system. The small amount of scatter in this dependency is due to variations in the temperature and radius of the WD---a fit for hot ({$T_{\rm eff}$} = 20,000\,K) WDs (top dotted line) and cold ({$T_{\rm eff}$} =  6,000\,K) WDs (bottom dotted line) is shown here, and an equation for this fit is given in the text. The magnitude of the WD is not a driving factor when determining the minimum detectable planet radius {\bf Right:}~Minimum detectable exoplanets of a given temperature-radius around WDs at four different distances ($2.7-13$\,pc). Earth-analogs can be detected around WDs within 6\,pc of Earth and hot, rocky planets with tenuous atmospheres (Mercury-analogs) within 10\,pc. Habitable-zone exoplanets orbiting WDs are detectable via IR excess out to $8-10$\,pc (e.g. Kepler-62e). Jupiters and Saturns are detectable with MRS out to $10-20$\,pc, but can more efficiently be detected with 21\,$\mathrm{\mu m}$ MIRI imaging. For both plots, we determine detectability based on the amount of IR excess measured from the exoplanet with 10\,hrs of observation in the JWST MIRI/MRS sub-band C channels (see Section~\ref{Mercury}). }
\label{minEarth}
\end{figure*}

For four WDs at various distances between 2.7-13\,pc, we determine the minimum-radius detectable ($5\sigma$) exoplanet as a function of exoplanet temperature. The results of this calculation are illustrated in Figure~\ref{minEarth}. From this figure, we see that Earth-analogs are detectable out to 6\,pc. Habitable-zone super-Earths (1.7R$_\oplus$), Jupiter, and Mercury are detectable around WD systems out to $\sim 10$\,pc. The left panel of Figure~\ref{minEarth} demonstrates that the minimum detectable size of a habitable-zone exoplanet in each system depends primarily on the distance to that system, rather than the properties of the host star.  This is because the IR excess technique relies primarily on collecting photons directly from the planet instead of performing a differential measurement as is done for exoplanets in eclipse. The small amount of scatter in the left panel of Figure~\ref{minEarth} is primarily due to differences in WD temperatures in our sample. WDs are all very similar in size, which minimizes what would otherwise be the most significant factor in determining the minimum detectable planet radius. We provide an equation to estimate the minimum detectable exoplanet radius (assuming 10\,hrs of MIRI/MRS observations) as a function of distance to the WD system and temperature of the white dwarf and exoplanet:
\begin{equation}
\frac{R_{min}}{R_\oplus}
= \left(0.16\left(\frac{{\rm d}_{\rm WD}}{1~{\rm parsec}}\right)+\left(\frac{T_{\rm eff}}{154,000 {\rm K}}\right) + 0.10\right){\left(\frac{T{_{\rm eq}}}{287{\rm K}}\right)^{-1.9}}
\end{equation}
where d$_{\rm WD}$ is the distance to the WD system in parsecs, $T_{\rm eff}$ is the temperature of the white dwarf in Kelvin and R$_{min}$ is the minimum detectable exoplanet radius in Earth radii. {The equation accurately estimates R$_{min}$ (to a precision of $\pm 25\%$) over the following range: $150$\,K$ < T{_{\rm eq}}$$ <$$ 700$\,K, $4000$\,K$ < T_{\rm eff}$$ <$$ 25,000$\,K and d$<$$15$\,pc.}

Since we are directly measuring the unresolved flux from the exoplanet, this detection technique is mostly insensitive to viewing angle. For the observations to be completely insensitive to inclination, at least one of the following must be true:
\begin{enumerate}
    \item We must observe the time-averaged flux (i.e.~a full orbit/phase-curve) in which case tidal locking would have no bearing on the exoplanet's mean flux from the system. For reference, an exoplanet in the habitable zone of our WD sample requires $\lesssim20$~hrs to complete an orbit for 69\% of our sample.
    \item Heat redistribution in the atmosphere of the exoplanet is sufficient such that the night-side of the planet is warm enough to allow for detection.  
\end{enumerate}
As long as one of these two conditions is true, then the statement that the technique is independent of inclination holds. Even if neither assumption is correct, the probability that the system is inclined near edge-on and that we happen to observe only the dark side of the planet is small. However, one might argue that for this reason, it is beneficial to observe hot WD systems (where case 1 does not hold) at several epochs to ensure that inclination/epoch does not prohibit exoplanet detection. If this is done, then IR excess can be used to detect exoplanets in {\it all} WD systems where the instrument is sufficiently sensitive.

It had been noted in the literature that JWST is capable of detecting biosignatures in the atmospheres of Earth-like planets transiting WDs \citep{2013MNRAS.432L..11L}. More recent calculations show that the biosignature pair of $O_3$+$CH_4$ for an edge-on Earth-like planet transiting WD 1856 ($d$ = 25pc, {$T_{\rm eff}$} = 4700K) are detectable (5$\sigma$) with 50 hours of observation \citep{2020ApJ...901L...1K}. However, the transit probability is low, and the transit method can only detect 1\% of habitable-zone Earths in WD systems \citep{2020ApJ...901L...1K}. % because the detection technique requires that the system inclination is edge-on. 
Since the IR excess method is able to detect nearly 100\% of habitable-zone Earths (in systems that are sufficiently nearby), the probability of detection is dependent only on the occurrence rate of exoplanets. {If a sufficient number of systems are observed, this technique could be used to constrain the occurrence rate of habitable-zone terrestrial planets around WDs. Constraints on occurrence rates are essential to understanding terrestrial planet formation and migration in post main-sequence stars.} Further, it has been proposed that WDs maintain a relatively stable habitable zone for billions of years \citep{2009ApJ...700L..30B,2011ApJ...731L..31A,2012ApJ...757L..15F,2013AsBio..13..279B,2013MNRAS.432..500N}. Measurements of their occurrence rates and nature would allow us to determine if they contribute significantly to the number of habitable worlds.

As a side note, we will briefly mention that these observations would also be sensitive to tidally-heated exomoons \citep{2013ApJ...769...98P,2021PSJ.....2..119R}, which would be dynamically stable around wide-separation directly-imaged gas-giant exoplanets. These ``Super-Ios", should  occupy the bottom right parameter space (hot, but small radii) of Figure~\ref{minEarth}.

\subsubsection{Noise Sources and Challenges of Detection}\label{sec:step1}

For habitable-zone Earth-like exoplanets and most rocky planets with tenuous atmospheres (those with temperatures $\lesssim 650K$), there is very little IR excess in channel 1 (6.4–7.5~$\mathrm{\mu m}$). Therefore, channel 1 is ideal for use as a reference channel. Specifically, to make these measurements, we will need to fit a WD model to our measurements at short wavelengths (where we do not anticipate contamination from an exoplanet). This model can then be subtracted off and the remaining excess flux from the source at longer wavelengths can be attributed to the presence of an exoplanet. Because the IR excess from terrestrial planets is likely to be small ($0.1-1\%$) this in practice will be challenging. Fortunately, there are several analysis techniques that can be implemented to facilitate recovery of the planet signal. Beyond that, for most terrestrial planets simultaneous detection of the phase curve and/or planetary atmosphere will provide several avenues for detecting (and confirming) the planet without solely relying on IR excess measurements. However, before we go into detail about atmosphere and phase curves, we will describe how to make precision IR excess measurements and the challenges we are likely to face.

First, as shown by our models in Section~\ref{sec:sample}, we expect the spectrum of a WD to be a featureless Rayleigh-Jeans tail. {To determine if there is infrared excess, we can first normalize our measurement and model at a shorter IR wavelength, and then determine if there is a relative excess in the measured data indicating a possible exoplanet. To accurately determine the WD temperature and make a very precise relative measurement requires building up a spectrum across the MIRI IR range ($5-22\mu m$). This can be done by supplementing the} primary MRS C sub-band observations with short observations in the MRS A and B sub-bands. The A, B and C sub-bands overlap which allows the construction of  a continuous SED of the WD, ensuring accurate calibration between channels within the sub-bands. Further, we can also obtain spectroscopy of the WD at shorter IR wavelengths with NIRSpec to ensure our WD model accurately captures the physics and measured SED of the WD. Relative flux calibration between sub-bands on the detector at a few hundred parts per million will be required to make these measurements. While the instrument has been designed to achieve these levels of precision, initial testing and observations with JWST will be necessary to confirm they are indeed achievable.

In addition to the instrumental systematics at the few hundred ppm level, we need also be concerned about IR excess from the WD itself. Although most WDs are featureless in the IR, some of these objects are known to deviate from a blackbody. To detect IR excess from a terrestrial exoplanet requires modeling the SED of a WD beyond $\sim10\mathrm{\mu m}$ with $\sim100ppm$ precision based on the measured WD SED at shorter wavelengths.

DA WDs are used as photometric standards because of their stable and predictable SED \citep{2019ApJS..241...20N,2020MNRAS.491.3613G}. {HST DA WD calibration} measurements are precise to 1\% and are in excellent agreement with models at that level of precision {\citep{2014PASP..126..711B}.} We will need to test if models are capable of predicting the SED at the higher precisions required for our proposed detection method, as previous studies have not acquired sufficiently precise measurements to test this. While most WDs are quiescent, there are some exceptions that are known to be variable above the 0.1\% level and should be avoided when searching for terrestrial planets via IR excess or phase curve modulations (see Section~\ref{sec:step2}). For example, in our sample, WD 1647+591 is a ZZ Ceti pulsator \citep{1983ApJ...271..744K} with a 0.6\% amplitude and a period of 109 seconds. WDs 1748+708, 0912+536, and 1900+705 have extreme magnetic fields ($\sim$100 MG), and in this regime existing models cannot reach a 0.1\% precision. Another potential issue with magnetic WDs (including the ones with smaller magnetic fields) is that they can have starspots (analogous to Sun spots) that induce some variability \citep{2005MNRAS.357..333B,2011ApJ...734...17V,2013MNRAS.433.1599L,2015ApJ...814L..31K}. While these exceptions preclude detection of terrestrial planets via IR excess and phase curve modulations, it should not preclude our ability to detect gas giants (see Section~\ref{gasGiants}) for which we only need $\sim$1\% photometric precision due to the large signal ($\gtrsim$10\% IR excess) from gas giant exoplanets at 21~$\mathrm{\mu m}$.

Even if we cannot predict the SED of the WD to the required precision, or if the system is contaminated with IR excess from a debris disk, we may be able to use exoplanet atmospheric features to differentiate between exoplanet IR excess, debris disks, and small deviations in the WD's SED. Specifically, where a WD SED deviates from a blackbody, these deviations tend to be very broadband -- absent of spectral features. We also expect about 3-4\% of WDs to host debris disks \citep{2012ApJ...760...26B,2015MNRAS.449..574R,2019MNRAS.489.3990R}. If a debris disk is present in the system, the derived area and disk temperature will help differentiate between planet and debris disk. If the debris disk is interior to the planet's orbit and is hotter than the exoplanet, this could interfere with the calibration from the shortest wavelength channel. For planets orbiting near the Roche limit, debris interior to the planet's orbit may be common where tidal forces are disintegrating the planet (e.g. \citealt{2015Natur.526..546V}). However, with sufficient spectral coverage, it may be possible to fit the SED of the disk and separate that component from the WD and exoplanet. An easier method for detecting exoplanets that possess atmospheres, would be detection of the $CO_2$ feature in MRS sub-band C channel 3. The presence of such a feature would best be explained by the presence of an exoplanet, since we anticipate featureless WD spectra and no $CO_2$ absorption from debris disk (see Section~\ref{sec:step3} for more discussion of spectral signature detection). Therefore, the presence of an atmosphere on the exoplanet may aid in our detection. We anticipate that this method may produce some false positives due to these factors, so any detections will need confirmation with phase curve observations or spectroscopy (see sections \ref{sec:step2} and \ref{sec:step3}). 

It is worth noting that IR excess can only be detected for small planets if the planet is warm. For cooler WDs, this means that the planet must be very close (within a few Roche radii) of the star to allow for detection. For hot WDs, such as Sirius B, we can detect exoplanets at much larger separations. For example, an exoplanet with an equilibrium temperature of 300K orbits at 0.14 AU around Sirius B (a hot, {$T_{\rm eff}$} = 25,967~K, WD), but at only 0.01 AU around WD 0046+051 (a much cooler, {$T_{\rm eff}$} = 6106~K WD). Therefore, this method (IR excess detection) is able to detect planets at a much larger range of separations around WDs that are hotter.

\subsection{Step 2: Confirming Candidates with Phase Curve Observations}\label{sec:step2}

As has previously been noted by \cite{2014ApJ...793L..43L}, the phase curves
of rocky worlds with no atmospheres orbiting WDs are detectable
with JWST. The observation time required to detect warm terrestrial exoplanets via IR excess of nearby WDs with JWST is generally comparable to the planet's orbital period. 
For example, the orbital period of a 300~K terrestrial planet orbiting WD 0046+051 is 10.3 hours. This implies that full orbital phase curves will simultaneously be measurable for close-in exoplanets in most cases. In these cases (especially for cool WDs where warm exoplanets must be on a very short orbit), {JWST could} measure a complete phase curve of warm exoplanets during initial detection/observation (no follow up required). WD 0046+051 has a typical temperature ({$T_{\rm eff}$} = 6106~K) for WDs in the solar neighborhood. Of the 34 WDs in our sample, 27 have temperatures between 4000-9000~K and 7 are between 9000-27,000 K. For the extremely hot WDs in the solar neighborhood (those with $T{_{\rm eff}}>$\,20,000K), full phase curves would only be measurable for extremely hot exoplanets.  Sparsely-sampled phase curves would still be possible. For Sirius B, a $\sim 1000$~K exoplanet takes about 10 hours to complete one orbit, whereas a $\sim 300$~K exoplanet would take $\sim 10$~days. Therefore, habitable-zone exoplanet phase curves are only measurable for the cooler WD systems (without additional follow-up observations) assuming $\sim 10$~hours are spent monitoring each system during initial observations.

{For an exoplanet similar in temperature to Mercury (day side of 700\,K), we would expect amplitude variations of 3-30 ppt for an exoplanet ranging between the radius of Mercury and Earth, respectively, This is detectable with JWST with a SNR of 10\,$\sigma$ (Mercury-sized planet) or 68\,$\sigma$ (Earth-sized planet).} This detection technique has been explored for WDs at visible wavelengths, but it should be much more sensitive in the mid-IR, where the planet-to-WD flux ratio is more favorable. The Kepler Space Telescope demonstrated time-domain stability of WDs at the $10-100$ part-per-million level for many WDs \citep{2015MNRAS.447.1749M, 2017MNRAS.468.1946H}, and much larger (part-per-thousand) amplitudes are expected from warm terrestrial exoplanets in the mid-IR. When measuring phase curves, it will be possible to monitor the change in flux in one band rather than measuring the IR flux relative to another spectral band. The complex calibrations of the spectrum required to detect IR excess discussed in Section~\ref{sec:detect} will not be necessary. Further, we need not predict the SED of the WD in the IR since we need only make a relative time-domain measurement.

\begin{figure*}
\centering
\includegraphics[width=1\textwidth]{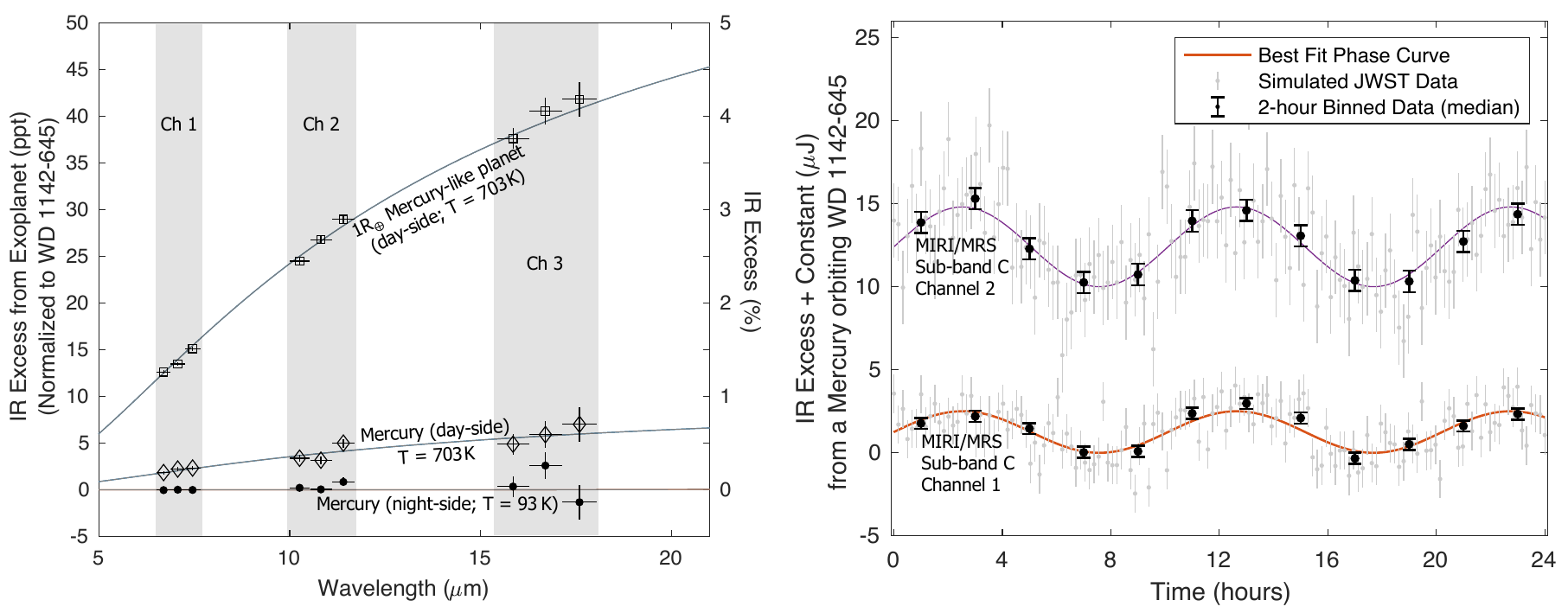}
\caption{{\bf Left:} IR excess in parts per thousand (left axis) or percent (right axis) from an exoplanet in the WD 1142-645 system (4.6\,pc) with Mercury's temperature distribution in three of the MIRI/MRS Sub-band C channels. IR Excess is plotted for a $T{_{\rm eq}}$ = 93~K (night-side), 1R$_{\rm Mercury}$ and 1R$_\oplus$ exoplanet (bottom line) and a $T{_{\rm eq}}$ = 703~K (day-side), 1R$_{\rm Mercury}$ and 1R$_\oplus$ exoplanets (middle and top lines, respectively). There is no detectable IR excess from the night-side of the exoplanet. Simulated, binned JWST MIRI/MRS data are shown assuming 10 hours of observation time. The spectral data for all three channels are collected simultaneously with the MRS instrument.  {\bf Right:} Simulated MIRI/MRS phase curve (24\,hrs of observation) for the 1R$_{\rm Mercury}$ exoplanet (day-side temperature of 703K) in the WD 1142-645 system. The Mercury-analog orbits at 0.01~AU (same day and night temperatures as left figure) and an orbital period of 10.1~hrs. The simulated data points (light gray) are 10-minute integrations with MRS, and the black points are binned (median) to 2 hour intervals. The IR excess and the phase curve from a Mercury-like exoplanet is detectable in this system.}
\label{MercuryPhaseCurve}
\end{figure*}

In short, unlike the IR excess detection approach, phase curves will be most sensitive to short-orbit ($<1$ day) exoplanets, but it will likely also be easier to implement as it is less sensitive to instrumental and astrophysical systematic errors. One way to mitigate this would be to break up observations of hot WDs into multiple observations (sparsely-sampled phase curves; \citealt{2016ApJ...824...27K}) to allow for detection of longer-period exoplanets via phase curve monitoring. These observations could still be combined to allow for IR excess detection as well.

Phase curve precisions with Spitzer were demonstrated with 40\,ppm precision \citep{2019Natur.573...87K} and measurements with JWST should be possible at the 8\,ppm level in the IR \citep{2016ApJ...832L..12K, Matsuo_2019,2021AJ....161..115S,2021PSJ.....2..140M}.
As shown in Section~\ref{sec:detect}, we expect signals $>1ppt$ from Earth-sized exoplanets with MIRI/MRS observations. While phase curve observations of main-sequence stars, especially M-dwarfs, must account for stellar variability, WDs will be temporally stable. Our limiting noise source is likely to be thermal background and photon noise. MIRI instrument noise (e.g., 1/f noise, interpixel sensitivity variations) may also contribute.

To illustrate terrestrial exoplanet phase curve detection, we model the MRS phase curves for a Mercury-analog orbiting WD 1142-645 spectral binned in three (of the four) channels, assuming an edge-on orientation. Figure~\ref{MercuryPhaseCurve} IR excess as a function of wavelength (left) and phase curve (right) from an exoplanet with Mercury's temperature distribution.

If initial MRS observations only detect IR excess, follow-up phase curve observations should be possible to confirm exoplanet candidates. Specifically, based on the measured equilibrium temperature of exoplanet candidates identified via only IR excess, it should be possible to constrain the orbital period of the planet. {Although, the constraints may be weak if there are a large number of degeneracies in the models based on atmospheric composition, planetary radius, and planetary photospheric/surface temperature or clouds.} However, as long as the planet has a measurable day-night temperature difference, either further continuous or intermittent (for longer period exoplanets) follow-up observations with JWST should allow for construction of a phase curve monitoring to confirm the exoplanet assuming there is a measurable day-night temperature difference and the system is not inclined near face-on.

\subsection{Step 3: Is there Life? -- Detecting Biosignatures}\label{sec:step3}

If an exoplanet is detected with an equilibrium temperature corresponding to the habitable zone {using the methods previously described} and the presence of an atmosphere is suspected, a search for biosignatures is surprisingly possible.% and comparatively quick. 

Here, we {will explore the detectability} of biosignatures (via methods described in sections \ref{sec:step1} and \ref{sec:step2}) for WD terrestrial exoplanets within $10$\,pc {(16 systems)} with JWST/MIRI using $5-36$ hours of integration time. 

The following process was used to calculate the sensitivity of LRS to ozone and methane. Note that in most cases, although an $N_2O$ absorption feature falls within the LRS spectral band (at 7.7\,$\mathrm{\mu}$m), the feature is much weaker than the $O_3$, $CH_4$ and $H_2O$ features, and will typically not be detectable.
We uploaded all the modeled WD spectra from our sample into the JWST/ETC. We used the LRS slit mode for these calculations. Here, since the goal is only detection of spectral features, the use of a slit is acceptable and necessary to minimize thermal background. We varied the total integration time to obtain sufficient SNR for biosignature detection. In all cases, the full subarray is read out in fast readout mode.

Similar to our previous MRS calculations, we used the ``lineplot\_sn.fits" file, which gives the SNR of the WD as a function of wavelength. We then binned the ETC SNR of the LRS measurements to the spectral width of the molecular feature and used this to determine the precision of our measurement. We determined the amount of signal in a molecular band by calculating the deficit of flux due to a molecular feature compared to what we would expect from a featureless blackbody (i.e. a planet with no atmosphere). 

\begin{figure*}
\centering
\includegraphics[width=1\textwidth]{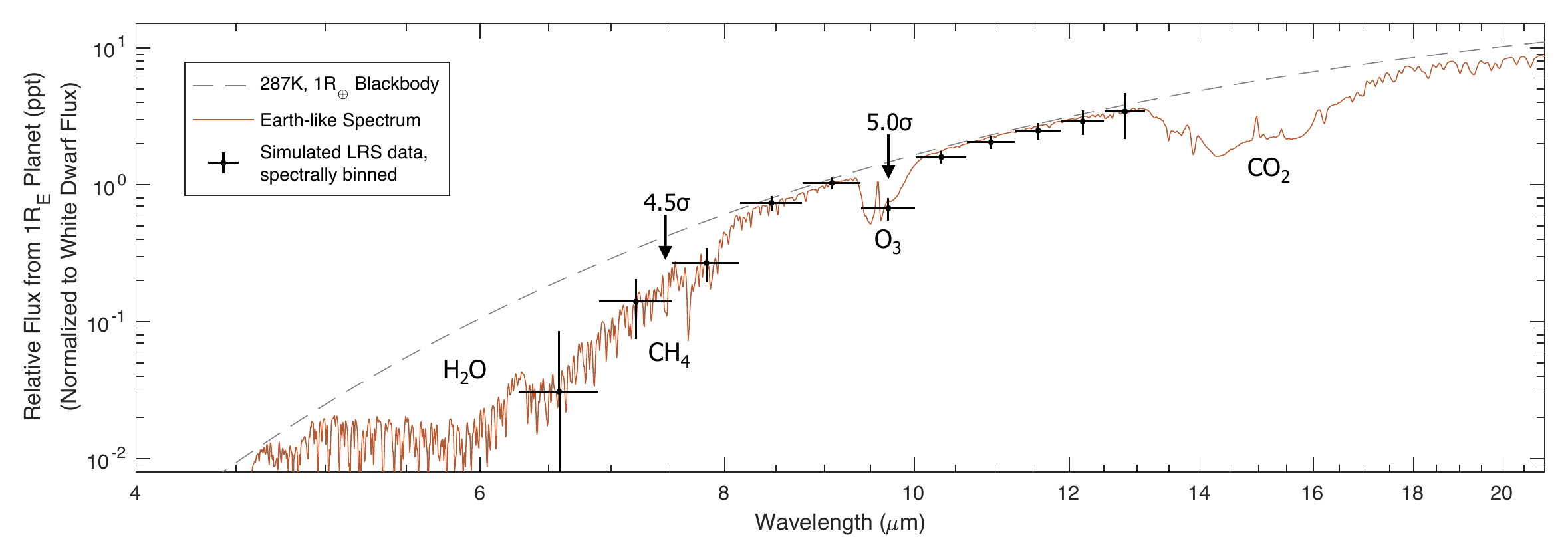}
\caption{Flux excess from an Earth analog orbiting WD 0046+051 vs. wavelength. The blackbody spectrum of a $T{_{\rm eq}}$=287~K, 1R$_\oplus$ (gray dashed line), the spectrum of an Earth-analog (red line) and simulated, binned  MIRI LRS measurements (black crosses). The biosignature pair $O_3$+$CH_4$ is detectable ($5\sigma$) in 25 hours for Earth-analogs in this WD system.}
\label{Biosignatures}
\end{figure*}

Using this method, we illustrate the detectability of the biosignature pair ($O_3$ and $CH_4$) on an Earth analog orbiting WD 0046+051. The results for 25 hours of observation time with LRS are shown in Figure~\ref{Biosignatures}. Here we show that the biosignature pair is detectable at the $>5\sigma$ level within 25 hours. We note that biosignature detection only requires about 5$\times$ the amount of time as planet detection. Specifically, this system would require 5 hours of MRS time for exoplanet IR excess detection, and 25 hours for LRS detection of the biosignature pair.

Earth's atmosphere has prominent methane and ozone absorption features in the IR at $7-8~\mathrm{\mu m}$ and $9-10~\mathrm{\mu m}$, respectively. The combined detection of these two molecules are a biosignature (i.e., indicative of life) \citep{2018AsBio..18..630M, 2020ApJ...894L...6K}. Here, we calculate the detectability of this molecular pair in the exoplanet atmospheres of our WD sample using MIRI/LRS. LRS is the best MIRI option for detecting biosignatures because the methane and ozone features fall within its 5-13 $\mathrm{\mu m}$ bandpass. However, the IR flux excess from these features is small ($0.1-1$ ppt). While we do expect sufficient sensitivity with JWST to detect these spectral features in this bandpass, detection of broad IR excess (from a featureless exoplanet with no atmosphere -- i.e. a blackbody) at this level of precision would likely not be possible. This is because for atmospheric features, and specifically these biosignatures, we are able to reference the flux in the molecular absorption band to the flux just outside the feature. For a featureless blackbody planet with no atmosphere, this is not possible due to the lack of a calibration or reference. Hence atmospheric features, and specifically Earth-like biosignatures are required in many cases to make an rocky exoplanet detectable with LRS. 

Although exoplanets should have spectral features even if they do not have biosignatures, the only other spectral features of Earth's atmosphere that fall in the LRS band are $H_2O$ and $N_2O$. While Earth's atmosphere contains all four of these species, Venus and Mars do not contain any molecular features at a detectable level in the LRS band. However, a hypothetical Earth-sized planet with substantial CH$_4$ in it's atmosphere -- e.g. a warm exo-Titan or an Archean Earth -- would have much stronger CH$_4$ absorption at $\sim$7.7 $\mathrm{\mu}$m than the Modern Earth atmosphere, which may be detectable using these observations. Note that the IR excess at the shorter (LRS) wavelengths is so small (unless the planet is extremely hot) it cannot be detected in broadband, but these narrow atmospheric absorption features (from methane and ozone) should be detectable in the spectrum. For that reason, it is best to use MRS for exoplanet IR excess detection and LRS for atmospheric and biosphere characterization.

%%%%%%%%%%%%%%%%%%%%%%%%%%%%%%%%%%%%%%%%%%%%%%%%%%%%%%%%%%%%%%%%%%%%%%
\section{Discussion}\label{sec:summary}

\subsection{Example JWST Programs to Search for WD Exoplanets via IR Excess}
\subsubsection{A Census of Gas-Giants}

A JWST/MIRI broadband 21\,$\mathrm{\mu}$m imaging survey of the 34 nearby WD systems for cold Jupiters requires only a couple hours of observation per target and, therefore, could be completed using a substantial mid-size JWST program. 
Such a survey would allow detection of much colder objects than previous surveys and would be sensitive to unresolved and directly imaged cold Jupiters at almost all separations (spanning the Roche limit out to hundreds to 1000 AU, depending on the distance to the WD - the field of view of MIRI is quite large; 74\farcs0 × 113\farcs0). 
For the nearest, young systems, our detection limits would probe down to Saturn mass exoplanets around nearby, young systems. For the oldest systems, at further distances ($\approx$13\,pc), the survey would still be sensitive to Jupiter analogs at almost all orbital separations (limited only by the field of view of MIRI). This survey would have an advantage over previous endeavors because (1) JWST is extremely sensitive which allows for a survey of many systems relatively quickly, (2) unlike previous surveys which have typically operated at shorter wavelengths (4.5\,$\mathrm{\mu}$m), cloudy exoplanets would be as easy to detect as those with clear atmospheres, (3) we can search for exoplanets at all orbital separations, and (4) JWST, even with short integration times, is sensitive to much smaller planets than previous surveys. 
Cold gas giant exoplanets directly imaged at 21\,$\mathrm{\mu m}$ could be confirmed via proper motion measurements, providing a second epoch of observations to increase confidence that the detection is not a background star.

A survey of these systems could constrain the occurrence rate of gas giant planets at all orbital separations with occurrence-rate precisions within a few percent. The WDs we consider here
have an average progenitor mass of 2\,M$_\odot$ \citep{2018ApJ...866...21C}. The
frequency of $>$1 AU gas giant planets around Sun-like stars is $>$20\% \citep{2021ApJS..255...14F} and the giant planet frequency around 2\,M$_\odot$ stars is 5$\times$ higher
than around 1\,M$_\odot$ \citep{2015A&A...574A.116R}. {Naively, one could thus expect gas giant planets to be ubiquitous around WDs. However, several uncertainties remain, including the survival rate of planets at the final stages of stellar evolution.}
If the occurrence rate of gas giant planets around WDs is similar to the occurrence rate around A-stars, then we would expect such a survey to detect ${\approx}10$ gas giant planets. An absence of detections would indicate that it is uncommon for gas giant planets, even at moderate to large separations, to survive the death of their host star. 
In contrast, if gas giants do out live their hosts, or if there is a planet re-genesis, it is not unreasonable that a mid-size JWST program could more than triple the number of known gas giant planets orbiting WDs. Detections or non-detections could provide insight into the formation mechanism of WD gas giants.

\subsubsection{A Search for Terrestrial Worlds}
A small JWST program could be used to test the MIRI/MRS terrestrial planet detection technique proposed in Section~\ref{sec:detect} and would aim to demonstrate detectability of IR excess, phase curves, and exoplanets with CO$_2$-dominated atmospheres. The two nearest solitary WDs (Wolf 28 and GJ 440, $d$ = 4.3\,pc and 4.6\,pc, respectively) are ideal targets. Observations with MIRI/MRS for $\approx$10 hours each will be sufficient to detect IR excess for HZ planets as small as $0.8-0.9$\,R$_\oplus$, phase curves from HZ exoplanets and Mercury-analogs, and provide sensitivity to molecular detection of CO$_2$-dominated atmospheres. If a habitable-zone planet is detected during observations, rapid (20-30\,hrs) follow-up observations with LRS could be used to search for signatures of life.  

\subsubsection{Finding Biosignatures in the Nearest Systems}
The two nearest WDs, Sirius B and Procyon B (2.7\,pc and 3.5\,pc, respectively) are both companions to bright, main sequence stars. For these systems, the bright star may contaminate MIRI/MRS observations because the MIRI/MRS field of view is comparable to the separation of the binaries. MRS observations may be feasible for Sirius B, where the binary separation is larger, but would be challenging in both cases. For these systems, if the primary goal is to find life, the best approach may be to use MIRI/LRS observations (in slit mode, which should block all light from the nearby bright star) to search for biosignatures in the spectrum rather than first trying to search for exoplanets with MRS. Because these systems are so close, biosignatures can be detected relatively quickly with LRS and a small JWST/MIRI/LRS program could be used to search for life around these WDs. However, the LRS observations would not be sensitive to all terrestrial worlds---only those with detectable levels of methane, water, ozone or nitrous oxide in their atmospheres---unless phase curves measurements with ppt precisions could be extracted despite the $0\farcs51$ slit. The Sirius system is quite young ($<250$\,Myr), and it is unclear if life could have had time to evolve, but Procyon B (1.4\,Gyr cooling age or 2.7\,Gyr total main sequence + WD age; \citealt{2015ApJ...813..106B}) is more ideal for such a study.

\subsection{A Comparison with Astrometric Detections}

Gaia is expected to detect a multitude of exoplanets, including a handful of planets orbiting WDs \cite{2015ASPC..493..455S,2022arXiv220602505S}. In this section, we briefly explore the detectability of WD exoplanets with Gaia astrometric measurements. Our goal is to compare the parameter space (e.g. planet size and orbital separation) of exoplanets that are detectable with the two methods (Gaia astrometry vs. JWST IR Excess/direct imaging).

\begin{figure}
\centering
\includegraphics[width=0.48\textwidth]{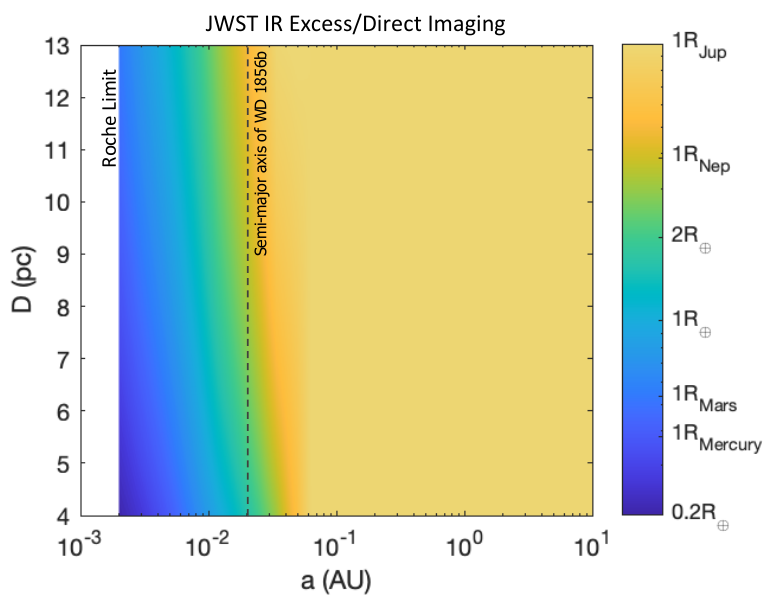}
\caption{The detectability of exoplanets orbiting nearby WDs with JWST using IR excess and direct imaging. Within the nearest 13\,pc, terrestrial planets at separations of $<$0.02\,AU are detectable around their host WD. Gas giant planets are detectable at all separations within the field of view of the MIRI instrument. The semi-major axis of the WD gas giant planet WD 1856b (black dashed line) is shown for reference. For this calculation, we assume the planets are blackbodies and use the detection method described in Section~\ref{Mercury}.}
\label{GAIAvIRExcess2}
\end{figure}

Figure~\ref{GAIAvIRExcess2} shows the exoplanet radii and separations that are detectable around a typical ({$T_{\rm eff}$} = 6000K) WD in our solar-neighborhood sample as a function of distance to the system. To this parameter space, we calculate the equilibrium temperature of a planet orbiting at a given distance from the star (assuming a uniform albedo of 0.3 and a planet with only blackbody emission). For that temperature, we then determine the minimum detectable planet radius with MRS sub-band C observations (using the same configurations described in Section~\ref{sec:detect}). We do this calculation at several distances between 4-13\,pc and then interpolate between. We find terrestrial exoplanets are detectable for separations $\lesssim$0.02\,AU (which happens to correspond to the semi-major axis of WD 1856b). Beyond that, gas giants are detectable at all separations out to the edge of the detector $10-100$\,AU for MRS, which has only a $7\farcs \times 8\farcs$ field of view. The mass limit of detectable gas giants is dependent on both the separation from the host star and the age of each individual WD system. We do not account for WD age in this calculation, but note that planetary-mass objects (between 1M$_{\rm Saturn}-10$M$_{\rm Jup}$, depending on the system's age) are detectable at all separations in each system. Exoplanets between $1-10$M$_{\rm Jup}$ are all represented on this plot with a 1$R_{\rm Jup}$ radius.

\begin{figure*}
\centering
\includegraphics[width=0.95\textwidth]{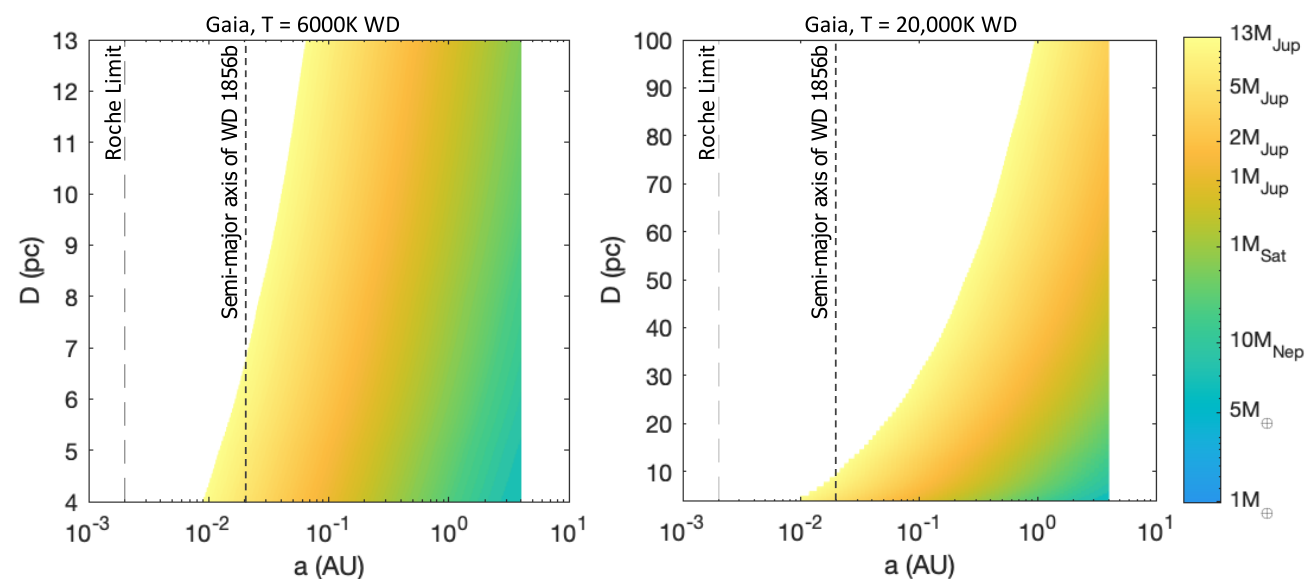}
\caption{The detectability of exoplanets orbiting nearby WDs with Gaia using astrometry on cold ({$T_{\rm eff}$} = 6000~K, left) and hot ({$T_{\rm eff}$} = 20,000~K, right) WDs. Gas giant exoplanets at separations of 0.05-5\,AU are detectable around nearby ($<$100\,pc) WDs with Gaia. Gaia is sensitive to sub-Saturn mass gas giants for systems within 15\,pc. For this calculation, we use the Gaia astrometric precisions from \citealt{2014ApJ...797...14P} (Table 2) and equation \ref{equ:astrometric-signature} to estimate precisions based on the 10-yr Gaia mission.}
\label{GAIAvIRExcess}
\end{figure*}

For comparison, we calculate the detectability of WD exoplanets with Gaia astrometry (see Figure~\ref{GAIAvIRExcess}). Note that we show the detectability of exoplanets around WDs out to 100\,pc for Gaia since Gaia is an all-sky survey, whereas a small JWST survey for exoplanets would likely be limited to only the nearest WDs (and therefore we only show detectability out to 13\,pc). For this calculation, we use the Gaia astrometric precisions from \cite{2014ApJ...797...14P} (Table 2). To estimate precisions for a 10-yr mission, we divide the provide precisions (given for the 5-yr mission) by a $\sqrt{2}$. We calculate the astrometric signature, $\alpha$, of the exoplanet using the equation given in \cite{2014ApJ...797...14P}, 
\begin{equation}
\label{equ:astrometric-signature}
\alpha = \left(\frac{M_{\rm p}}{M_{\rm WD}}\right) \left(\frac{a_{\rm p}}{\rm 1~AU}\right) \left(\frac{d}{\rm 1\,pc}\right)^{-1} \hspace{-5pt} {\rm arcsec} \ ,
\end{equation}
where $d$ is the distance, and $M_{\rm p}$ and $M_{\rm WD}$ the mass of the planet and WD, respectively. 
We again use a typical cool ({$T_{\rm eff}$} = 6000~K) WD from our sample, which corresponds to a V = 14.5~mag star at 10\,pc. The 5~$\sigma$ exoplanet detection limit is shown for the cool WDs on the left in Figure~\ref{GAIAvIRExcess}.
Astrometric detections with Gaia work best at $0.5-4$ AU, with possible detection down to Neptune-mass exoplanets for the nearest WD systems.

Unlike JWST observations, which must be conducted on a star-by-star basis, Gaia will cover the entire sky. Therefore, in addition to looking at the detectability of exoplanets around the nearby WDs in our sample, we also look at the detectability of hotter (20,000~K) WDs out to 100\,pc (Figure~\ref{GAIAvIRExcess}, right).

JWST observations are able to detect terrestrial exoplanets that are close to the host star, whereas Gaia is more sensitive to longer period gas giants. But there is some overlap between the exoplanets that are detectable with Gaia and JWST. Most of the exoplanets that are detectable with Gaia would be characterizable with follow-up JWST IR Excess detection or direct imaging observations.

\subsection{Dead Stars Might be a Good Place to Search for Life}
Biosignature detection and the development of exoplanet characterization techniques have traditionally been focused on characterizing exoplanets orbiting main sequence stars. Indeed, an observatory capable of characterizing exoplanets analogous to Earth was a primary recommendation of the Astro2020 Decadal Survey\footnote{\url{nap.nationalacademies.org/initiative/decadal-survey-on-astronomy-and-astrophysics-2020-astro2020}}. There has also been a recent push to characterize HZ exoplanets transiting M-dwarfs, which are more amenable to JWST observations than FGK stars due to the favorable transit depths. However, biosignature detection for exoplanets in the most favorable systems, TRAPPIST-1, would still require hundreds of hours of JWST observations \citep{2019AJ....158...27L,2021MNRAS.505.3562L}. Further, previous studies have shown WDs provide a stable habitable zone for billions of years, possibly conducive to the formation of life  (see discussion and references at the end of section \ref{RRlimits}). Despite the field’s focus on detecting life on exoplanets orbiting main sequence stars, we have demonstrated in this manuscript that there may be HZs where JWST can much more easily detect biosignatures. 

With the future launch of a direct-imaging Flagship mission observatories, we could be able to characterize and detect life on habitable zone worlds around main sequence stars. However, the habitable zone of WDs lies too close to the host star to be accessible to coronagraphic imaging in the foreseeable future. Therefore, if our goal is to search for life around all nearby star systems, the technique presented here offers the first possible method for biosignature detection around all nearby WD systems. Unlike future coronagraphic or star-shade imaging with large space telescopes, this biosignature detection method is accessible to us now. There are no planned upcoming IR space observatories with sensitivities comparable to JWST, so observations with the MIRI instrument may be our only chance to search for life around nearby WDs. 

{Although the technique is in principle capable of detecting biosignatures, it is important to keep in mind that there is only a small number of systems for which this biosignature detection method is possible, and we do not know (i.) if Earth-like planets around WD exist, and (ii.) if they do exist, if any of them host exoplanets with detectable biosignatures.}

\section{Conclusions}\label{sec:conc}

In this work we have proposed a new technique for detecting white dwarf exoplanets and characterizing their atmospheres with the {\it James Webb Space Telescope}. Our main results can be summarized as follows.

\begin{enumerate}
    \item JWST can detect the infrared excess from unresolved, cold ($>150$\,K) gas-giant exoplanets orbiting white dwarfs within 15\,pc using 2\,hrs of MIRI broadband 21\,$\mathrm{\mu}$m observations per system. A mid-size JWST program leveraging this technique is capable of performing an efficient census of nearby white dwarf gas-giant exoplanets.
    \item JWST can detect the infrared excess from unresolved, temperate or hot ($>250$\,K) terrestrial exoplanets, including Earth and Mercury analogs, orbiting white dwarfs within 10\,pc using 10\,hrs of MIRI medium-resolution spectroscopy. This is the only technique capable of detecting most (non-transiting) terrestrial worlds around nearby white dwarfs.
    \item Follow-up MIRI low-resolution spectroscopy can be used to search for biosignatures on HZ white dwarf terrestrial worlds. This is the only known technique for detecting biosignatures on non-transiting white dwarf exoplanets in the solar neighborhood.
    \item While both JWST (via IR excess \& direct imaging) and Gaia (via astrometry) are capable of detecting nearby gas-giant planets orbiting white dwarfs, only JWST is sensitive to terrestrial worlds.
\end{enumerate}

In our solar neighborhood ($<$10 pc), there are 74 AFGK stars, 283 M-dwarfs, 21 WDs, and 50 (currently known) sub-stellar objects. If we place an Earth-analog in the HZ around each of these 428 objects, JWST can detect the biosignatures of an Earth-analog orbiting {the nearest six WDs (within 7\,pc)} most readily, requiring $\lesssim 25$\,hrs in each system\footnote{Assuming none of these HZ worlds transit.}.
If an abundance of life exists in this obscure location we are likely to detect biosignatures on these worlds with JWST in the near future---{\it if we choose to look for it}---long before we have observatories capable of characterizing Earth-analogs orbiting main sequence stars.

%% Acknowledgements %%
\section*{Acknowledgements}
The authors thank the reviewer, Ren\'{e} Heller, for a thorough and helpful review of this manuscript. We thank Darren L. DePoy for his helpful conversations and suggestions. MAL acknowledges support from the George P. and Cynthia Woods Mitchell Institute for Fundamental Physics and Astronomy at Texas A\&M University. As part of the CHAMPs (Consortium on Habitability and Atmospheres of M-dwarf Planets) team, KBS and JLY acknowledge support by the National Aeronautics and Space Administration (NASA) under Grant No. 80NSSC21K0905 issued through the Interdisciplinary Consortia for Astrobiology Research (ICAR) program. SB is a Banting Postdoctoral Fellow and a CITA National Fellow, supported by the Natural Sciences and Engineering Research Council of Canada (NSERC).

This research has made use of the NASA Exoplanet Archive, which is operated by the California Institute of Technology, under contract with the National Aeronautics and Space Administration under the Exoplanet Exploration Program.

\section*{Data Availability}
All data are incorporated into the article.

\bibliography{main.bib}
\bibliographystyle{mnras}
%\end{document}

%%%%%%%%%%%%%%%%%%%%%%%%%%%%%%%%%%%%%%%%%%%%%%%%%%

% Don't change these lines
\bsp	% typesetting comment
\label{lastpage}
\end{document}